\documentclass[twocolumn,showkeys,showpacs,preprintnumbers,prd,superscriptaddress,nofootinbib]{revtex4-1}
\usepackage{graphicx,epsf,bm,amsmath,amsfonts,amssymb,epstopdf,natbib,hyperref,color,verbatim,multirow,bm,appendix,array,diagbox}
\hypersetup{colorlinks=true,urlcolor=blue,citecolor=blue,linkcolor=blue,menucolor=blue,anchorcolor=blue,filecolor=blue}

\newcolumntype{?}{!{\vrule width 3pt}}
\date{\today}

\definecolor{darkgreen}{RGB}{0,100,0}
\usepackage{orcidlink}
\newcommand{\x}{{\bf x}}
\def\msun{\ M_\odot}
\newcommand{\be}{\begin{equation}}
\newcommand{\ee}{\end{equation}}
\newcommand{\bea}{\begin{equation}\begin{aligned}}
\newcommand{\eea}{\end{aligned}\end{equation}}

\def\dd{{\rm d}}

\def\lsim{\mathrel{\rlap{\lower4pt\hbox{\hskip0.5pt$\sim$}}\raise1pt\hbox{$<$}}}
\def\gsim{\mathrel{\rlap{\lower4pt\hbox{\hskip0.5pt$\sim$}}\raise1pt\hbox{$>$}}}

\begin{document}
\title{Hunting Dark Matter with the Einstein Telescope}
 
\author{A.J.~Iovino \orcidlink{0000-0002-8531-5962}}
\affiliation{New York University, Abu Dhabi, PO Box 129188 Saadiyat Island, Abu Dhabi, UAE}
 
\author{M.~Maggiore \orcidlink{0000-0001-7348-047X}}
\affiliation{Department of Theoretical Physics and Gravitational Wave Science Center,\\
24 quai E. Ansermet, CH-1211 Geneva 4, Switzerland}
 
\author{N.~Muttoni \orcidlink{0000-0002-4214-2344}}
\affiliation{Department of Theoretical Physics and Gravitational Wave Science Center,\\
24 quai E. Ansermet, CH-1211 Geneva 4, Switzerland}
 
\author{A.~Riotto \orcidlink{0000-0001-6948-0856}}
\affiliation{Department of Theoretical Physics and Gravitational Wave Science Center,\\
24 quai E. Ansermet, CH-1211 Geneva 4, Switzerland}
 
\begin{abstract}
Too light primordial black holes evaporate and are therefore strongly constrained by various bounds, e.g.\ Cosmic Microwave Background distortion. However, if they are formed strongly clustered, the corresponding haloes may collapse into heavier black holes which may form the entirety of the dark matter of the universe. The indirect signal of such a scenario is the production of a flat stochastic background of gravitational waves which is detectable by the Einstein Telescope.
\end{abstract}
 
\maketitle

\section{Introduction}
 
Primordial Black Holes (PBHs)\,\cite{Carr:1974nx,Carr:1975qj} have recently attracted significant attention, as they may constitute the entirety of Dark Matter (DM) in the so-called asteroidal mass range defined as
\begin{equation}
\label{range}
    10^{-16}\,M_\odot \lesssim M_{\rm PBH} \lesssim 10^{-11}\,M_\odot\,,
\end{equation}
and generate gravitational-wave (GW) signals potentially detectable by current and future experiments (see e.g.\ Ref.\,\cite{Carr:2020gox,Green:2020jor,Escriva:2022duf,Carr:2026hot} for  recent reviews).
 
Among the various formation mechanisms, the standard scenario involves the collapse of large-amplitude curvature perturbations generated during inflation. The characteristic scale $k$ at which the enhancement of these perturbations occurs determines the typical PBH mass via the horizon-mass relation
\begin{equation}
    M_{\rm PBH}\equiv \mathcal{O}(0.1)\,M_H\simeq M_\odot \left(\frac{k}{10^{6}\,{\rm Mpc}^{-1}}\right)^{-2}\,:
\end{equation}
a perturbation of comoving scale $k^{-1}$ re-enters the horizon when $aH\sim k$, and since the Hubble mass scales as $M_H\sim(aH)^{-3}\rho\propto k^{-2}$ during radiation domination, the PBH mass inherits this scaling. At the same time, the same scale sets the frequency of the associated scalar-induced gravitational-wave (SIGW) background, since the GWs are produced at horizon re-entry and their present-day frequency is redshifted from that moment, giving%
\footnote{We note that this relation is approximate: PBHs forming from a single characteristic scale $k$ are produced with a range of masses due to the spread in the amplitude of the collapsing perturbations, and the relation above should be understood as giving the peak of the resulting mass function.}
(for a recent review see Ref.\,\cite{Domenech:2021ztg})
\begin{equation}
    f \simeq 1.6\,{\rm nHz}
    \left(\frac{k}{10^6\,{\rm Mpc}^{-1}}\right)
    \simeq 1.6\,{\rm nHz}
    \left(\frac{M_{\rm PBH}}{M_\odot}\right)^{-1/2}.
\end{equation}
This relation establishes a direct link between PBH mass and GW frequency, enabling the exploration of PBH formation across a wide mass range through different GW observatories.
 
On the one hand, the space-based interferometer LISA~\cite{LISACosmologyWorkingGroup:2022jok} will probe the mHz frequency band, corresponding to asteroid-mass PBHs, which can account for the totality of DM (see Refs.~\cite{Bartolo:2018evs,Bartolo:2018rku,LISACosmologyWorkingGroup:2024hsc,LISACosmologyWorkingGroup:2025vdz,Iovino:2025cdy} for works on the topic). On the other hand, the Einstein Telescope (ET)\,\cite{Punturo:2010zz,ET:2019dnz,ET:2025xjr} will be sensitive to GW signals in the $\mathcal{O}(10)\,$Hz band, corresponding to the formation of much lighter PBHs, $M_{\rm PBH}\sim10^{-21}\,M_\odot$, which would have already evaporated by today, leading to strong constraints on their abundance because of, e.g., the distortion in the CMB. As a consequence, it is often argued that ET is unable to probe the hypothesis that PBHs make up the entirety of dark matter, except perhaps indirectly.
 
In this paper we show that in fact ET may give us a hint about the DM. Suppose PBHs are strongly clustered, beyond the Poisson expectation, at the time of formation~\cite{Chisholm:2005vm,Tada:2015noa,Ballesteros:2018swv,Belotsky:2018wph,Desjacques:2018wuu,Ali-Haimoud:2018dau,Inman:2019wvr,Suyama:2019cst,Matsubara:2019qzv,MoradinezhadDizgah:2019wjf,Young:2019gfc,Atal:2020igj,DeLuca:2020jug,DeLuca:2021hcf,Gorton:2022fyb,Petac:2022rio,DeLuca:2022bjs,Animali:2024jiz,Auclair:2024jwj,Crescimbeni:2025ywm}. If so, PBHs are formed in close proximity, and their subsequent gravitational interactions can lead to the collapse of the entire cluster into a heavier PBH, even during the radiation-dominated era, thereby evading the bounds associated with evaporated PBHs. This mechanism is commonly referred to as \emph{clusterogenesis}\,\cite{DeLuca:2022bjs}. Very light PBHs (which potentially may evaporate and so, for instance, distort the CMB spectrum) if strongly clustered may not only form a heavier and harmless PBH, but their mass can easily fall in the range where they can form the entirety of the DM. The intriguing bonus is that the stochastic SIGW spectrum generated at formation of the light PBHs at second order falls in the range detectable by ET, so that the latter may help in the hunt for DM and understanding its nature.
 
The paper is organized as follows. In Section~II we show why the correlation function of the compaction function and of the curvature perturbation are linked to each other, and using basic concepts of probability conservation, we provide the expression for the correlation function in the case of a broad spectrum. In Section~III we determine the condition for which \emph{clusterogenesis} happens. The direct consequences on the gravitational-wave signal produced, with a particular focus on the ET frequency band, are discussed in Section~IV. We conclude in Section~V.
 
\section{The PBH clustering}
 
\subsection{Setting the stage}
 
Before launching ourselves into the details, some necessary technical remarks.
The key starting object is the curvature perturbation $\zeta(\bf x)$ on superhorizon scales appearing in the metric in the comoving uniform-energy density gauge
\be
{\rm d}s^2=-{\rm d}t^2+a^2(t)\,e^{2\zeta(\bf x)}{\rm d}{\bf x}^2\,,
\ee
where $a(t)$ is the scale factor in terms of the cosmic time. Cosmological perturbations may gravitationally collapse to form a PBH depending upon the amplitude measured at the peak of the compaction function\,\cite{Shibata:1999zs}, defined to be the mass excess compared to the background value in a given radius\,\cite{Musco:2018rwt}
\be
C(\x)=2\frac{M(\x,t)-M_{\rm b}(\x,t)}{R(\x,t)}\,,
\ee
where $M(\x,t)$ is the Misner-Sharp mass, with $M_{\rm b}(\x,t)$ its background value. The Misner-Sharp mass gives the mass within a sphere of areal radius $R(\x,t)=a(t)r\,e^{\zeta(\bf x)}$ with spherical coordinate radius $r$, centred around position $\x$ and evaluated at time $t$. The compaction directly measures the overabundance of mass in a region and is therefore better suited than the curvature perturbation for determining when a region collapses. It can be written in terms of the density contrast as
\be
C(\x)=\frac{2}{R(\x,t)}\int {\rm d}^3\x\,\rho_{\rm b}\,\delta(\x,t)\,,
\ee
where $\rho_{\rm b}$ is the background energy density and on superhorizon scales the density contrast is related to the curvature perturbation in real space by the nonlinear relation\,\cite{Harada:2015yda}
\be
\delta(\x,t)=-\frac{4}{9}\frac{1}{a^2H^2}\,e^{-2\zeta(\bf x)}\!\left(\nabla^2\zeta(\x)+\frac{1}{2}\left({\mathbf\nabla}\zeta(\x)\right)^2\right).
\ee
Assuming spherical symmetry and defining $\zeta'={\rm d}\zeta/{\rm d}r$, the compaction function becomes
\begin{equation}
\label{c}
C(r)=C_1(r)-\frac{3}{8}C_1^2(r),\quad
C_1(r)=-\frac{4}{3}\,r\zeta'(r)\,,
\end{equation}
where
\begin{equation}
C(r)=\frac{8\pi}{R(r,t)}\rho_{\rm b}\int_0^R{\rm d}\widetilde{R}\,\widetilde{R}^2(r,t)\,\delta(r,t)\,.
\end{equation}
Notice that the compaction function is half of the density perturbation averaged inside the radius $r$ at horizon entry.
 
To characterize the PBH two-point correlation function $\xi(r)$ (or, simply, correlation function) at any comoving separation $r=|{\bf r}|$, we can use the overdensity of discrete PBH centers at position ${{\bf r}}_i$ (eventually smoothed over a sphere with a radius equal to the Hubble radius at the moment the perturbations re-enter the horizon)
\begin{equation}\label{eq:Cluster1}
\delta_{\rm PBH}({\bf r})=\frac{1}{\overline{n}_{\rm PBH}}\sum_i\delta_D({\bf r}-{\bf r}_i)-1\,,
\end{equation}
where $\delta_D({\bf x})$ is the three-dimensional Dirac distribution, $n_{\rm PBH}$ is the average comoving number density of PBHs, and $i$ runs over the initial positions of PBHs. The corresponding two-point correlation function must take the general form
\bea\label{def}
\big\langle\delta_{\rm PBH}({\bf x})\delta_{\rm PBH}(0)\big\rangle
&= \frac{1}{\overline{n}_{\rm PBH}}\delta_D({\bf x})-1\nonumber\\
&\quad+\frac{1}{\overline{n}_{\rm PBH}^2}\Bigl\langle\sum_{i\neq j}\delta_D({\bf x}-{\bf x}_i)\delta_D({\bf x}_j)\Bigr\rangle\nonumber\\
&= \frac{1}{\overline{n}_{\rm PBH}}\delta_D({\bf x})+\xi({\bf x})\,,
\eea
where\,\cite{DeLuca:2022bjs}
\begin{equation}\label{eq:n_PBH}
\overline{n}_{\text{\rm PBH}}\simeq 30\,f_{\text{\rm PBH}}\left(\frac{M_{\rm PBH}}{M_\odot}\right)^{-1}\mathrm{kpc}^{-3}.
\end{equation}
The first term in the final step of Eq.~(\ref{def}) is the unavoidable Poisson contribution coming from the fact that PBHs are discrete objects. In the scenario where PBHs are formed at the collapse of sizeable overdensities at re-entry within the comoving horizon ${\cal H}^{-1}$, we have
\be
\overline{n}_{\text{\rm PBH}}\sim P_1{\cal H}^3\,,
\ee
where $P_1$ is the probability to form a PBH inside the comoving horizon.
 
\subsection{A peak of the compaction function for a peak of the curvature perturbation}
 
Suppose now that there is a peak in the curvature perturbation $\zeta(\x)$ with a given peak value $\zeta(0)$ and profile $\zeta(r)$ away from the center, which we arbitrarily set at the origin of the coordinates. It is easy to convince oneself that to each peak of the curvature perturbation corresponds a peak of the compaction function. Performing a rotation of the coordinate axes to be aligned with the principal axes of length $\lambda_i$ ($i=1,2,3$) of the constant-curvature-perturbation ellipsoid and Taylor-expanding to second order one obtains\,\cite{Bardeen:1985tr}
\be
\zeta(r)\simeq\zeta(0)-\frac{1}{2}\sum_{i=1}^3\lambda_i x_i^2\,,\quad r^2=\sum_{i=1}^3 x_i^2\,.
\ee
Large-amplitude peaks are approximately spherical with a common positive $\lambda$~\cite{1986ApJ...304...15B}. Therefore
\be
C_1(r)\simeq\frac{4}{3}\lambda r^2\,,
\ee
from which we deduce that, around the point ${\bf x}=0$, the function $C_1(r)$ increases. Since at large radii it must decrease and vanish, there must be a value, let us call it $r_m$, at which $C_1(r)$ has a maximum. Now, since
\be
\label{max}
C'(r_m)=C'_1(r_m)\left(1-\frac{3}{4}C_1(r_m)\right)=0\,,
\ee
the extremum of the compaction function $C(r)$ coincides with the extremum of its linear counterpart $C_1(r)$. Furthermore,
\be
C''(r_m)=C''_1(r_m)\left(1-\frac{3}{4}C_1(r_m)\right)\,,
\ee
and the maximum of the compaction function $C(r)$ coincides with the maximum of its linear counterpart $C_1(r)$ as long as $C_1(r_m)<4/3$ (that is, we consider type I PBHs), which we safely assume from now on.
 
To locate the thresholds of the compaction function it is therefore enough to locate those of $C_1$. We stress, however, that in full generality a peak in $\zeta$ need not coincide with a peak in the compaction: the compaction $C(\mathbf{x},r)$ is a smoothed quantity evaluated at the scale $r$, and for a broad power spectrum the correlation coefficient between $C$ and $\zeta$ satisfies $\gamma_{c\zeta}\to0$ as the spectral width $\Delta\to\infty$~\cite{Young:2022phe}. The physically relevant perturbations for PBH formation are, however, rare, high-amplitude overdensities in the exponential tail of the distribution. For such perturbations, peaks in the overdensity $\delta$, and hence in the compaction $C$, are spatially co-located with peaks of $\zeta$ that are sufficiently ``spiky'', i.e.\ that satisfy~\cite{DeLuca:2019qsy}
\be
\label{eq:spiky}
-\nabla^2\zeta(\mathbf{x}_{\rm pk})
> \frac{9\,a^2 H^2}{4}\,e^{2\zeta(\mathbf{x}_{\rm pk})}\,\delta^c_{\rm pk}\,.
\ee
This condition is automatically more easily satisfied for rarer, higher-threshold peaks. In our scenario, PBH formation occurs at the UV end of the broad power spectrum, at scale $k_{\rm max}$, where the relevant perturbations lie precisely in the exponential tail and the high-peak limit ensures approximate spherical symmetry~\cite{Bardeen:1985tr}. Under these conditions the argument of Ref.~\cite{DeLuca:2019qsy} applies and the co-location of peaks in $\zeta$ and in $C$ is justified. The correlation function of PBHs is therefore the correlation function of the fully non-Gaussian curvature perturbation, which we now study.
 
\subsection{The full non-Gaussian two-point correlation}
 
Consider now a generic fully non-Gaussian curvature perturbation, whose probability has a non-Gaussian tail at large values. It is important to remark at this stage that from now on we will be dealing with classical quantities in the sense that we focus on the perturbations whose wavelengths are larger than the Hubble radius at the end of inflation. Without loss of generality, we can express $\zeta$ as a functional of a Gaussian counterpart $\zeta_{\rm G}$,
\be
\zeta({\bf x})=F\left[\zeta_{\rm g}({\bf x})\right]\,,
\ee
with joint probability $P[\zeta({\bf x}_1),\cdots,\zeta({\bf x}_n)]$. As a concrete example, we can mention the relation
\be
\label{eq:NG}
\zeta=-\mu\ln\!\left(1-\frac{\zeta_{{\rm g}}}{\mu}\right),
\ee
which arises naturally in standard models such as the Ultra-slow-roll inflationary scenario~\cite{Atal:2019cdz} and the curvaton mechanism~\cite{Sasaki:2006kq,Pi:2022ysn}. In the first case, $\mu$ is typically related to the slope of the inflationary power spectrum of $\zeta_{{\rm g}}$ beyond its peak. In models in agreement with CMB observations we have $1.5\lsim\mu\lsim5$~\cite{Atal:2018neu,Karam:2022nym,Frosina:2023nxu}. For curvaton models, $\mu$ is intricately determined by the time of the curvaton's decay into radiation and the shape of the power spectrum~\cite{Lyth:2002my,Enqvist:2005pg,Lyth:2006gd,Ando:2018nge,Ferrante:2023bgz,Allegrini:2026jqt}, with typical values of $\mathcal{O}(0.1)$. Expanding Eq.~\eqref{eq:NG} to second order, we find that $f_{\text{\rm NL}}=5/(6\mu)$. We remark however that our arguments are not restricted to the expression~(\ref{eq:NG}).
 
Now, in threshold statistics, the correlation of the thresholds of the fully non-Gaussian curvature perturbation is
\bea
1+\xi(r)&=\frac{P_2}{P^2_1}\,,\nonumber\\
P_1&=\int_{\zeta>\zeta_{\rm c}}\dd\zeta\,P[\zeta]\,,\nonumber\\
P_2&=\int_{\zeta(0)>\zeta_{\rm c}}\!\!\dd\zeta(0)\int_{\zeta(r)>\zeta_{\rm c}}\!\!\dd\zeta(r)\,P[\zeta(0),\zeta(r)]\,,
\eea
where $\zeta_{\rm c}$ is the threshold. The key point is the conservation of probability
\bea
P[\zeta]\,\dd\zeta&=P[\zeta_{{\rm g}}]\,\dd\zeta_{{\rm g}}\,,\nonumber\\
P[\zeta(0),\zeta(r)]\,\dd\zeta(0)\,\dd\zeta(r)&=P[\zeta_{{\rm g}}(0),\zeta_{{\rm g}}(r)]\,\dd\zeta_{{\rm g}}(0)\,\dd\zeta_{{\rm g}}(r)\,,
\eea
where we assume that the functional $F$ is monotonically increasing in the physical domain, so that the map $\zeta_{\rm g}\mapsto\zeta$ is one-to-one and no sum over branches is required. For the specific example of Eq.~\eqref{eq:NG}, the map $\zeta=-\mu\ln(1-\zeta_{\rm g}/\mu)$ is indeed strictly monotonically increasing for $\zeta_{\rm g}<\mu$, which is the physical domain of the field.
 
From the last equivalence, we deduce that the correlation function of the fully non-Gaussian curvature perturbation can be derived from the Gaussian counterpart, provided that one appropriately shifts the threshold
\be
\zeta_{{\rm cg}}=F^{-1}\left[\zeta_{\rm c}\right].
\ee
For the example, from Eq.~(\ref{eq:NG}),
\be
\zeta_{{\rm cg}}=\mu\left(1-e^{-\zeta_{{\rm c}}/\mu}\right).
\ee
We conclude that we can calculate everything in terms of Gaussian probabilities,
\begin{align}
1+\xi(r)&=\frac{P_{2{\rm g}}}{P^2_{1{\rm g}}}\,,\nonumber\\
P_{1{\rm cg}}&=\int_{\nu_{{\rm g}}}^\infty\frac{\dd x}{\sqrt{2\pi}}e^{-x^2/2}
={\rm Erfc}\!\left(\frac{\nu_{{\rm cg}}}{\sqrt{2}}\right),\nonumber\\
P_{2{\rm cg}}&=\int_{\nu_{{\rm g}}}^\infty\frac{\dd x_1}{\sqrt{2\pi}}
\int_{\nu_{{\rm cg}}}^\infty\frac{\dd x_2}{\sqrt{2\pi}}\nonumber\\
&\quad\times\exp\!\left[-\frac{x_1^2+x_2^2-2\,w_g x_1 x_2}{2(1-w_g^2)}\right],
\end{align}
where
\begin{equation}
\nu_{{\rm cg}}=\frac{\zeta_{{\rm cg}}}{\sigma_{{\rm g}}}\,,\qquad
\sigma_{{\rm g}}^2=\int\frac{\dd k}{k}\,{\cal P}_{\zeta_{\rm g}}(k)\,,
\end{equation}
and
\be
w_{\rm g}(r)=\frac{\xi_{\rm g}(r)}{\sigma_{{\rm g}}^2}
\ee
is the (normalized) two-point correlation of the Gaussian curvature perturbation. A valid approximation for $\nu_{{\rm g}}\gg1$ has been found in Ref.~\cite{Ali-Haimoud:2018dau} to be
\be\label{eq:xi}
1+\xi(r)\simeq(1+w_{\rm g})\,
\frac{{\rm Erfc}\!\left(\sqrt{\frac{1-w_{\rm g}}{1+w_{\rm g}}}\frac{\nu_{{\rm cg}}}{\sqrt{2}}\right)}{{\rm Erfc}\!\left(\frac{\nu_{{\rm cg}}}{\sqrt{2}}\right)}\,.
\ee
An important remark is the following: the numerical value of the critical thresholds or of the shift from $\zeta_{\rm c}$ to $\zeta_{\rm cg}$ is not relevant. By the same argument of probability conservation, the PBH abundance for the fully non-Gaussian $\zeta$ can be derived from the one-point probability of $\zeta_{\rm g}$ for which only the value of $\nu_{\rm g}$ is needed. This, in turn, is fixed phenomenologically by requiring that the PBH abundance has a given value. This fixes the parameter $\nu$, independently from the value of the threshold $\zeta_{\rm cg}$.
 
Another interesting remark is the following. In the limit of large non-Gaussianities, $\mu\ll1$, the threshold of the non-linear curvature perturbation is in the non-Gaussian exponential tail of the probability when $\zeta_{\rm c}\gg1$, or $\zeta_{\rm cg}\simeq\mu\ll1$ from Eq.~(\ref{eq:NG}). This means that, when the threshold for the curvature perturbation is in the non-Gaussian exponential tail, the correlation becomes independent from the formation threshold: when expressed in terms of the Gaussian counterpart through the conservation of the probability, the Gaussian threshold $\zeta_{\rm cg}$ tends to zero. This simple reasoning confirms the result of Ref.\,\cite{Animali:2024jiz}: when the threshold of the non-linear $\zeta$ is in the exponential tail region of the probability, the correlation loses its dependence on the threshold $\zeta_c$ of the non-linear field. We stress that the correlation still depends on $\nu_{\rm cg}=\zeta_{\rm cg}/\sigma_{\rm g}$, but since $\zeta_{\rm cg}\simeq\mu$ saturates to a constant independent of $\zeta_c$, this is not a dependence on the non-linear threshold itself.
 
What values of $\nu_{\rm cg}$ should we use? The present abundance of DM in the form of PBHs per logarithmic mass interval $\mathrm{d}\ln M$ is given by
\begin{align}
\label{f}
f_{\rm PBH}(M)
&\equiv\frac{1}{\rho_{{\rm DM}}}\frac{{\rm d}\rho_{\rm PBH}}{{\rm d}\ln M}\nonumber\\
&\simeq\left(\frac{P_1}{6\cdot10^{-9}}\right)
\left(\frac{\gamma}{0.2}\right)^{1/2}
\left(\frac{106.75}{g_*}\right)^{1/4}
\left(\frac{M_\odot}{M}\right)^{1/2}\nonumber\\
&=\left(\frac{P_{1{\rm g}}}{6\cdot10^{-9}}\right)
\left(\frac{\gamma}{0.2}\right)^{1/2}
\left(\frac{106.75}{g_*}\right)^{1/4}
\left(\frac{M_\odot}{M}\right)^{1/2},
\end{align}
where $\gamma\lsim1$ is a parameter introduced to take into account the efficiency of the collapse and, for the masses of interest, the number of relativistic degrees of freedom $g_*$ can be taken to be the SM value $106.75$. In the second passage we have used the key information that $P_1=P_{1\rm g}$. For PBH masses of the order of $10^{-21}\,M_\odot$, potentially testable by ET, $P_1=P_{1{\rm g}}$ should be of the order $2\cdot10^{-19}$ to get $f_{\rm PBH}(10^{-21}\,M_\odot)\simeq1$, corresponding to $\nu_{\rm cg}\simeq9$.
 
Provided that the threshold is appropriately shifted, we have demonstrated that the correlation function of the full non-Gaussian curvature perturbation is identical to that of the Gaussian counterpart
\be
1+\xi_{>\nu_{\rm c}}(r)=\frac{P_{2{\rm g}}}{P^2_{1{\rm g}}}=1+\xi^{\rm g}_{>\nu_{\rm cg}}(r)\,,
\ee
where $\xi^{\rm g}_{>\nu_{\rm cg}}(r)$ is the correlation of the thresholds of the Gaussian curvature perturbation. Given a profile of $\zeta_{\rm g}(r)$ we can compute the corresponding value $r_m$ at which the compaction function $C_1$ has its maximum. For the profile, we take the averaged profile
\bea
\langle\zeta_{\rm g}(r)\rangle_{\zeta_{\rm g}(0)}&=\langle\zeta_{\rm g}(r)|\zeta_{\rm g}(0)\rangle\nonumber\\&=\frac{\langle\zeta_{\rm g}(r)\zeta_{\rm g}(0)\rangle}{\sigma_{\rm g}^2}\,\zeta_{\rm g}(0)\\&=w_{\rm g}(r)\,\zeta_{\rm g}(0)\,.
\eea
For the most optimistic case, in terms of clustering, of a broad spectrum
\begin{equation}\label{eq:PSbroad}
   {\cal P}_{\zeta_{\rm g}}(k)=A_{\rm b}\,\theta(k_{\rm max}-k)\,\theta(k-k_{\rm min})\,,
\end{equation}
one has
\begin{align}
w_{\rm bg}(r)&=\frac{1}{\ln(k_{\rm max}/k_{\rm min})}\left[{\rm Ci}(k_{\rm max}r)-{\rm sinc}(k_{\rm max}r)\right.\nonumber\\
&\quad\left.-(k_{\rm max}\rightarrow k_{\rm min})\right],\\[6pt]
C_{1{\rm b}}(r)&\simeq-\frac{4}{3}\frac{\zeta_{\rm g}(0)}{\ln(k_{\rm max}/k_{\rm min})}\left[{\rm sinc}(k_{\rm max}r)-{\rm sinc}(k_{\rm min}r)\right]\,,
\end{align}
with $k_{\rm max}r_m\simeq3.5$. As shown in Ref.~\cite{MoradinezhadDizgah:2019wjf}, in the case of a broad spectrum the PBH mass function is peaked at the mass formed when the comoving wavelength $k^{-1}_{\rm max}$ re-enters the comoving Hubble radius.
\begin{figure*}[t!]
\begin{center}
\includegraphics[width=.75\textwidth]{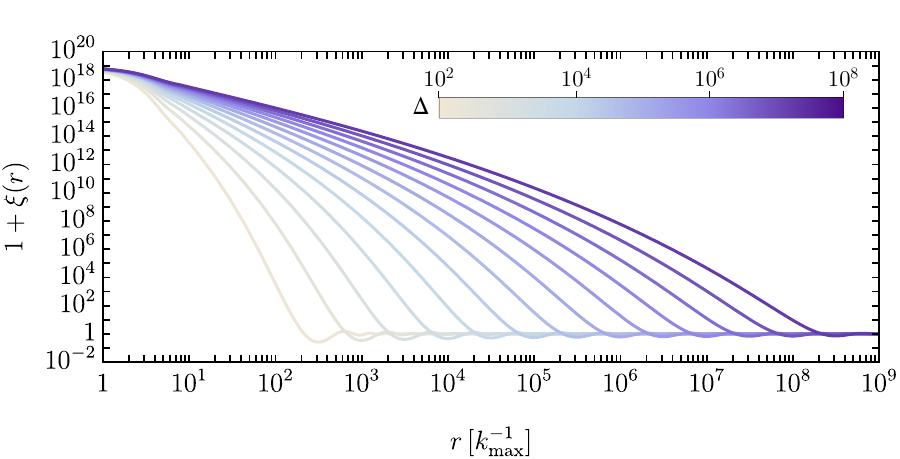}
\caption{\em The two-point correlation length $\xi(r)$ from Eq.\,\eqref{eq:xi} for a broad USR power spectrum from Eq.\,(\ref{eq:PSbroad}), fixing the benchmark value with $\nu_{\rm cg}=9$ and tuning the length $\Delta$.}
\label{fig:xiR}
\end{center}
\end{figure*}
 
The two-point correlation function from Eq.~\eqref{eq:xi}, evaluated at the shifted threshold $\nu_{\rm cg}=9$ which accounts for the full non-Gaussianity via the conservation of probability, is plotted in Fig.~\ref{fig:xiR}, where from now on we identify the length of the power spectrum as $k_{\rm max}\equiv10^\Delta k_{\rm min}$.
 
We caution, however, that the co-location of peaks does not imply a 1-to-1 quantitative correspondence between the threshold value of $C$ and the threshold value of $\zeta$: the nonlinear relation between the two fields introduces corrections that depend in general on the amount of non-Gaussianity\,\cite{DeLuca:2019qsy,young:2019yug}. These corrections do not affect the qualitative conclusions of the paper. Indeed the correlation function $\xi(r)$ depends on the non-Gaussian curvature perturbation only through the rescaled threshold $\nu_{\rm cg}=\zeta_{\rm cg}/\sigma_{\rm g}$, which is fixed by requiring $f_{\rm PBH}=1$, independently of the precise values of $\zeta_{c}$ or $C_{c}$ and independently of the shape of the non-Gaussianity. Any nonlinear correction to the threshold mapping is therefore automatically reabsorbed into the phenomenological determination of $\nu_{\rm cg}$; for reasonable amounts of non-Gaussianity this translates into a small shift in the amplitude $A_{\rm b}$ required to achieve $f_{\rm PBH}\sim1$, leaving the qualitative predictions for PBH clustering and the associated SIGW signal unchanged.
 
\section{The PBH clusterogenesis}
 
\subsection{The number of PBHs in the cluster}
 
At what minimum value of the radius should one evaluate the correlation function? In its original form, the thresholded regions are a continuous field and thus do not include any exclusion. This spatial exclusion arises because distinct PBHs cannot form arbitrarily close to each other. As a result, the conditional probability to find a PBH at a comoving distance $r$ from another PBH, which is proportional to $[1+\xi(r)]$, must vanish for some $r\lsim r_{\rm exc}$ and the correlation function of thresholded regions should go to $-1$ for comoving distances smaller than $r_{\rm exc}$.
 
Now, the physical meaning of $r_m$ is the following. Imagine having a peak in the Gaussian curvature $\zeta$. Given our assumptions, this has a spherical profile whose maximum value is at ${\bf x}=0$. The corresponding compaction function $C_1(r)$ has a maximum at a distance $r_m$ from the peak of the curvature perturbation. It is the spherical region of radius $r_m$ which will eventually collapse into a PBH upon horizon re-entry when $r_m={\cal H}^{-1}$, where ${\cal H}$ is the comoving horizon scale. From such considerations, we conclude that the correlation function must be evaluated at least at the distance
\be
r_{\rm exc}\gtrsim2r_m\,,
\ee
to avoid the superposition of two peak profiles overlapping.
 
In order to determine if PBHs are born clustered at formation, we should calculate the average number of PBHs in the comoving volume $V$ centered on a given PBH, which receives contributions both from the mean density and from the two-point correlation,
\be
\label{mean}
\langle N\rangle=\overline{n}_{\rm PBH}\int_V\!\dd^3r\,[1+\xi(r)]
=\overline{n}_{\rm PBH}V+\overline{n}_{\rm PBH}\int_V\!\dd^3r\,\xi(r)\,.
\ee
The first term is the Poisson expectation set by the mean abundance, while the second term captures the excess due to clustering. The mean count $\langle N\rangle$ significantly deviates from Poisson if the second contribution from the clustering is larger than the Poisson background $\overline{n}_{\rm PBH}V$. To quantify the importance of the clustering term $\bar{n}_{\rm PBH}\int_V\mathrm{d}^3r\,\xi(r)$, we will compute the characteristic (comoving) clustering length $r_{\rm cl}$ at which $\xi(r_{\rm cl})\simeq1$. However, the existence of a clustering length does not imply automatically that correlation is relevant. Indeed, from Eq.~(\ref{mean}), the Poisson contribution $\bar{n}_{\rm PBH}V$ to the average number of PBHs increases like the volume centered around a given PBH and therefore scales like $r^3$. On the other hand, the piece from the correlation function scales slower than $r^3$. Therefore, even if the two contributions become of the same order at $r_{\rm cl}$, one has to also estimate the average number of bubbles in a volume of size the correlation length, that is $\overline{n}_{\rm PBH}r_{\rm cl}^3$. Since $\overline{n}_{\rm PBH}\sim P_1{\cal H}^{3}=(P_1/r_m^3)$, all in all, the average number of PBHs in the cluster is
\bea
\langle N\rangle &\equiv
\frac{4\pi^2 P_1}{r_m^{3}}\int_{r_{\rm exc}}^{r_{\rm cl}}{\rm d}r\,r^2\,\xi(r) \\ &=4\pi^2 P_1\int_{2}^{r_{\rm cl}/r_m}{\rm d}x\,x^2\,\xi(x)\gsim1\,,
\eea
with $x\equiv r/r_m$.
 
\subsection{Clusterogenesis}
 
Once the cluster has formed, it may form a virialized halo even during the radiation phase. Indeed, it is usually believed that overdensities may not grow during the radiation phase because of the presence of the radiation pressure. However, this is true only at the linear level. If the overdensities are large enough, the self-gravity of these large non-linear fluctuations may become sizeable before the equality time, and produce, after they decouple from the Hubble flow, very dense clusters~\cite{Kolb:1994fi}.
 
Adopting for simplicity a spherical collapse model, which is justified since the clusters satisfying the Hoop conjecture are rare, high-amplitude overdensities expected to be approximately spherically symmetric in the high-peak limit~\cite{Bardeen:1985tr} (see also Refs.~\cite{Bamber:2025gxj,Delos:2024poq} for recent numerical studies of the collapse dynamics), the evolution equation for the parameter $R$, describing the deviation of the motion of each collapsing shell from the uniform Hubble flow of the background Friedmann universe reads\,\cite{Kolb:1994fi}
\be
x(1+x)\frac{\dd^2 R}{\dd x^2}+\left(1+\frac{3}{2}x\right)\frac{\dd R}{\dd x}
+\frac{1}{2}\left(\frac{1+\Phi}{R^2}-R\right)=0\,,
\ee
where
\be
\Phi=\frac{\delta n_{\rm PBH}}{\overline{n}_{\rm PBH}}\simeq\langle N\rangle\,,
\ee
we have assumed that all the DM is in PBHs, and $x=a/a_{\rm eq}$ in terms of the scale factor at the epoch of matter-radiation equality $a_{\rm eq}$. For small values of $x$, one may obtain analytically
\be
R\simeq1-\frac{\Phi x}{2}-\frac{\Phi^2 x^2}{8}+\mathcal{O}(x^3)\,.
\ee
The decoupling from the Hubble flow takes place at $a_{\rm cl}=a_{\rm eq}/\Phi$, well in the radiation phase. The physical radius of the cluster is\,\cite{DeLuca:2022bjs}
\bea\label{eq:Deluca1}
R_{\rm cl}&\simeq4\cdot10^{-5}
\langle N\rangle^{-1}\!\left(\frac{C}{200}\right)^{-1/3}\!
\left(\frac{r_{\rm cl}}{\rm kpc}\right){\rm kpc}\,,\\
M_{\rm cl}&\simeq 1.3\cdot10^2\langle N\rangle\left(\frac{r_{\rm cl}}{\rm kpc}\right)^3\msun
\eea
where $C=\mathcal{O}(1\text{--}10^2)$ parametrizes the amplitude of the final overdensity upon virialization\,\cite{Kolb:1993zz,Kolb:1994fi}.
\begin{figure*}[t!]
\begin{center}
\includegraphics[width=.75\textwidth]{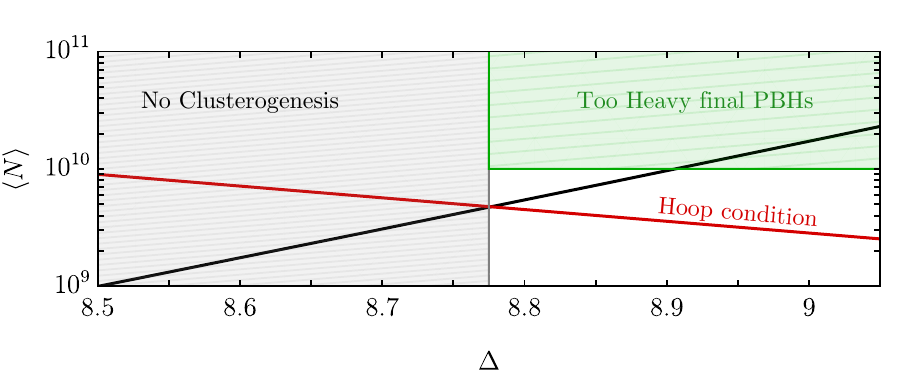}
\caption{\em  The average number of PBHs of masses $\sim10^{-21}\msun$ inside a volume of size $r_{\rm cl}\sim k^{-1}_{\rm min}$ for different power spectrum lengths $\Delta$. The red line sets the minimum number of PBHs in the cluster such as to satisfy the Hoop conjecture, i.e.\ Eq.\,(\ref{Eq:Hoop}). We set for simplicity $C=200$.}
\label{fig:N}
\end{center}
\end{figure*}
 
PBH clusters may collapse into a heavier BH if the final halo is at least as compact as a black hole of the same mass, thereby giving rise to a population of heavier PBHs. This condition is described by the Hoop conjecture\,\cite{Misner:1973prb}
\be
R_{\rm cl}<2G M_{\rm cl}\,,
\ee
or, combining Eqs.~(\ref{eq:Deluca1}) with the Hoop condition, we get
\begin{equation}
    \langle N\rangle>6\cdot10^4\left(\frac{C}{200}\right)^{-1/6}\left(\frac{r_{\rm cl}}{\textrm{kpc}}\right)^{-1}.
\end{equation}
In our case, $r_{\rm cl}\simeq k^{-1}_{\rm min}=10^{\Delta}k^{-1}_{\rm max}$, and the relation between the mass of the lightest PBHs and the scale $k_{\rm max}$ is given by\,\cite{Sasaki:2018dmp}
\begin{equation}
    M_{\rm PBH}\simeq2.5\cdot10^{6}\left(\frac{k_{\rm max}}{\textrm{kpc}^{-1}}\right)^{-2}M_\odot\,,
\end{equation}
with $k_{\rm max}\simeq5\cdot10^{13}\,\textrm{kpc}^{-1}$ in order to get $M_{\rm PBH}\sim10^{-21}\msun$%
\footnote{Notice that in this paper we are interested in asteroid-like PBH masses, much lighter than the ones discussed in Refs.~\cite{DeLuca:2022uvz,Crescimbeni:2025ywm}.}.
Setting for simplicity $C=200$, the Hoop conjecture can then be recast in terms of the width of the power spectrum $\Delta$ and the PBH mass as
\begin{equation}\label{Eq:Hoop}
    \langle N\rangle\gtrsim9.5\cdot10^7\cdot10^{-\Delta}\sqrt{\frac{M_\odot}{M_{\rm PBH}}}\,.
\end{equation}
As shown in Fig.\,\ref{fig:N}, satisfying the Hoop conjecture requires a rather extended power spectrum, with 
\begin{equation}\label{eq:min}
    \Delta\gtrsim8.8.
\end{equation} 
This sets a lower limit on the broadness of the power spectrum.
 
Another important requirement is that PBHs, which typically evaporate on a timescale\,\cite{Carr:2026hot}
\begin{equation}
    \tau_{\rm eva}\approx1.0\times10^{65}\left(\frac{M_{\mathrm{PBH}}}{M_\odot}\right)^3\,\text{yr}\,,
\end{equation}
do not evaporate before the characteristic free-fall time that sets the formation time of the heavier PBHs. An additional constraint arises from the dissipation time of the cluster due to repeated encounters among its constituents, which may lead to its disruption before collapse into heavier PBHs\,\cite{1987gady.book.....B}. This effect is unrelated to Hawking evaporation. We neglect it here since, as shown in Ref.\,\cite{DeLuca:2022bjs}, it is less stringent than the Hoop conjecture\,\cite{1987gady.book.....B}:
\begin{equation}
\label{freefall}
\tau_{\rm cl}=\sqrt{\frac{3\pi}{32G\rho_{\rm cl}}}
\simeq1.2\cdot10^4\langle N\rangle^{-2}\left(\frac{C}{200}\right)^{-1/2}\,\text{yr}\,.
\end{equation}
For $M_{\rm PBH}\sim10^{-21}\,M_\odot$ and $k_{\rm max}/k_{\rm min}\sim10^{8.8}$, which implies $\langle N\rangle\gtrsim4\cdot10^{9}$, one can easily verify that, for any value $C\in[1,200]$, the condition $\tau_{\rm cl}\ll\tau_{\rm eva}$ is satisfied. The same conclusion holds for spinning BHs, whose evaporation time is approximately a factor of two shorter than in the non-spinning case\,\cite{Arbey:2019jmj}.

An upper bound on the broadness of the power spectrum follows from the requirement that the final PBH population still accounts for the totality of the DM. In other words, the final mass must lie in the range where current observational constraints allow $f_{\rm PBH}=1$, namely~\cite{Carr:2026hot}
\begin{equation}\label{eq:window}
    10^{-16}\,M_\odot\lesssim M^{\rm final}_{\rm PBH}\lesssim10^{-11}\,M_\odot\,.
\end{equation}
Since the final mass is given by $M^{\rm final}_{\rm PBH}\simeq\langle N\rangle M_{\rm PBH}$, starting from $M_{\rm PBH}\sim10^{-21}\,M_\odot$ the condition~(\ref{eq:window}) requires
$
10^{5}\lesssim\langle N\rangle\lesssim10^{10}\,,$
which translates into an upper limit on the width of the power spectrum, $\Delta\lesssim8.9$. Combining it with the lower limit of Eq.~(\ref{eq:min}) for a successful clusterogenesis, we obtain
\begin{equation}
8.8\lesssim\Delta\lesssim8.9\,.
\end{equation}
Such a broad spectrum can be obtained in the context of curvaton models\,\cite{Ferrante:2023bgz} and within USR models with a prolonged constant-roll phase%
\footnote{At the moment there are no explicit USR potentials in the literature producing a nearly scale-invariant power spectrum large enough to have a successful clusterogenesis.}\,\cite{Leach:2001zf,Franciolini:2022pav}.
In the next subsection, we investigate the implications for detection prospects in current and future experiments.

We stress that the Hoop condition~(\ref{Eq:Hoop}) is a sufficient, not a necessary, criterion for the light PBHs to be absorbed into a heavier one. Clusters which do not satisfy it are nevertheless not a population of free-streaming light PBHs: decoupling from the Hubble flow occurs at $a_{\rm cl}=a_{\rm eq}/\Phi$ and only requires $\Phi\simeq\langle N\rangle\gg1$, so that any such overdensity still virialises on the free-fall time $\tau_{\rm cl}$ of Eq.~(\ref{freefall}) and becomes a gravitationally bound halo of PBHs. Such a halo subsequently undergoes gravothermal core collapse on a timescale $t_{\rm cc}\simeq15\,t_{\rm rh}$, with $t_{\rm rh}\simeq0.138\,\langle N\rangle\,\tau_{\rm cl}/\ln(0.4\langle N\rangle)$ the half-mass relaxation time\,\cite{1987gady.book.....B}, i.e. 
\begin{equation}\label{eq:tcc}
t_{\rm cc}\simeq\frac{2.5\cdot10^{4}}{\langle N\rangle\,\ln(0.4\langle N\rangle)}\left(\frac{C}{200}\right)^{-1/2}{\rm yr}\,,
\end{equation}
after which the seed formed in the core accretes the remaining members of the same bound halo. Requiring $t_{\rm cc}<\tau_{\rm eva}$ gives $\langle N\rangle\gtrsim70$ for $C=200$ (and $\langle N\rangle\gtrsim200$ in the most conservative case $C=1$), to be compared with $\langle N\rangle>4\cdot10^{9}$ from Eq.~(\ref{Eq:Hoop}). At our benchmark, $t_{\rm cc}\simeq7\,$s against $\tau_{\rm eva}\simeq100\,$yr. Light PBHs therefore end up in a heavier BH well before evaporating even in the sub-critical clusters, and the fraction which never does so is negligible.
\begin{figure*}[t!]
\centering
\includegraphics[width=0.75\textwidth]{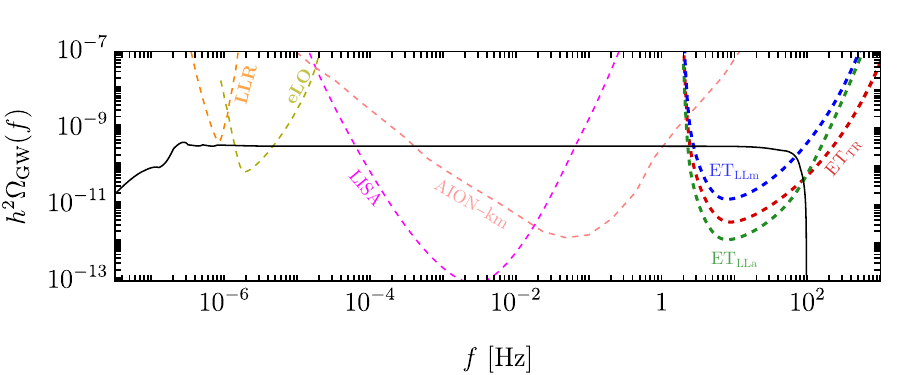}
\caption{\it Signal of second-order GWs associated with the broad power spectrum obtained for the benchmark case relevant in this work. We plot the power-law-integrated sensitivity curves (thick dashed lines) for three configurations of ${\rm ET}$: the triangular configuration (${\rm ET}_{\rm TR}$), the 2L-aligned configuration (${\rm ET}_{\rm LLa}$) and the 2L-misaligned (${\rm ET}_{\rm LLm}$) (in the sense discussed in the main text). We assume $1$~yr of observation and signal-to-noise ratio~$10$. We also plot the power-law-integrated sensitivity curves (dashed thin lines) of ${\rm LLR}$ and ${\rm eLO}$~\cite{Foster:2025nzf,Blas:2026xws}, ${\rm LISA}$~\cite{LISA:2022kgy} and ${\rm AION\text{-}km}$~\cite{Badurina:2019hst,Badurina:2021rgt,Abdalla:2024sst}.}
\label{fig:situation}
\end{figure*}

\section{Implication for detection with Einstein Telescope}
 
As already mentioned, PBH production is related to a signal of induced GWs. Such GWs are produced as a second-order effect by scalar perturbations which re-enter the horizon and collapse to form a PBH\,\cite{Tomita:1975kj,Matarrese:1993zf,Acquaviva:2002ud,Mollerach:2003nq,Ananda:2006af,Baumann:2007zm} (see also~\cite{Iovino:2025xkq} for a recent physical explanation of the origin of the signal). The spectral density of induced GWs associated to PBH production can be computed as\,\cite{Espinosa:2018eve}
\begin{widetext}
\begin{align}\label{eq:OmegaGW}
h^2\Omega_{{\rm GW}}(f)=\frac{c_{g}\,h^2\Omega_{r}}{36}
\int_{0}^{1/\sqrt{3}}\dd t\int_{1/\sqrt{3}}^{\infty}\dd s\,
\left[\frac{\left(t^{2}-1/3\right)\left(s^{2}-1/3\right)}{t^{2}-s^{2}}\right]^{2}
\left[\mathcal{I}^2_{c}(t,s)+\mathcal{I}_{s}^2(t,s)\right]
\mathcal{P}_{\zeta_{\rm g}}\!\left[\frac{k\sqrt{3}}{2}(s+t)\right]
\mathcal{P}_{\zeta_{\rm g}}\!\left[\frac{k\sqrt{3}}{2}(s-t)\right],
\end{align}
\end{widetext}
with
\begin{equation}
c_g\equiv\frac{g_*(M_H)}{g_{*}^0}
\left(\frac{g_{*S}^0}{g_{*S}(M_H)}\right)^{4/3}\approx0.4\,,
\end{equation}
($g_{*S}$ and $g_{*}$ being the effective degrees of freedom of thermal radiation and the superscript $^0$ standing for present-day values), $\Omega_r$ the current radiation density, and $\mathcal{I}_{c}(t,s)$, $\mathcal{I}_{s}(t,s)$ are transfer functions computed analytically assuming a perfect radiation cosmological background (see, for example, Ref.\,\cite{Espinosa:2018eve}).
 
Notice therefore that Eq.\,(\ref{eq:OmegaGW}) is strictly valid only during radiation domination. In principle, we should also consider the evolution of the relativistic degrees of freedom as the Electroweak phase transition takes place. We neglect this effect and just assume the Universe to be radiation dominated at all the scales relevant for GW production. We also neglect possible primordial non-Gaussian corrections in the computation of the scalar-induced GW signal (we refer to Refs.\,\cite{Cai:2018dkf,Cai:2018dig,Unal:2018yaa,Cai:2019elf,Hajkarim:2019nbx,Atal:2021jyo,Yuan:2020iwf,Domenech:2021and,Garcia-Saenz:2022tzu,Liu:2023ymk,Yuan:2023ofl,Li:2023xtl,Adshead:2021hnm,Perna:2024ehx,Inui:2024fgk,Iovino:2024tyg,Iovino:2024sgs} for a discussion about these effects).

Their size can be estimated as follows. Expanding Eq.\,(\ref{eq:NG}) one has $\zeta=\zeta_{\rm g}+\zeta^2_{\rm g}/2\mu+\cdots$, so that the parameter controlling the non-Gaussian corrections to the induced signal is not $f_{\rm NL}$ itself but $\sigma_{\rm g}/2\mu$: the GW background is sourced by the typical fluctuations, of amplitude $\sigma_{\rm g}$, and not by the rare peaks which collapse into PBHs. Since $\zeta_{\rm cg}\le\mu$, one has $\sigma_{\rm g}=\zeta_{\rm cg}/\nu_{\rm cg}\le\mu/\nu_{\rm cg}$ and therefore
\begin{equation}\label{eq:NGcorr}
\frac{\sigma_{\rm g}}{2\mu}\le\frac{1}{2\nu_{\rm cg}}\simeq\frac{1}{18}\,,
\end{equation}
independently of the value of $\mu$. At leading order the non-Gaussian contribution adds a positive-definite term to the source, so that the effect is an enhancement of the signal, bounded in relative amplitude by a few times $10^{-3}$ and equal to $\sim10^{-3}$ for the benchmark adopted below. We refer to Ref.\,\cite{Perna:2024ehx} for the fully non-Gaussian computation of the induced spectrum. Since the effect goes in the direction of increasing the amplitude, the detectability discussed below is conservative in this respect.

In Fig.\,\ref{fig:situation} we report the Gaussian SIGW signal for the relevant cases discussed in the previous section, i.e.\ a broad power spectrum (Eq.\,\ref{eq:PSbroad}) with $k_{\rm max}=5\cdot10^{13}\,\textrm{kpc}^{-1}$ and $k_{\rm min}\equiv10^{-\Delta}k_{\rm max}$ with $\Delta=8.8$. The amplitude $A_{\rm b}$ is not a free parameter: once the model of non-Gaussianity is specified, it is fixed by the requirement $f_{\rm PBH}\simeq1$. Indeed, from Eq.\,(\ref{eq:PSbroad}) one has $\sigma^2_{\rm g}=A_{\rm b}\,\Delta\ln10$, so that
\begin{equation}\label{eq:Ab}
A_{\rm b}=\frac{1}{\Delta\ln10}\left(\frac{\zeta_{\rm cg}}{\nu_{\rm cg}}\right)^{2}\,,
\end{equation}
with $\nu_{\rm cg}\simeq9$. The threshold $\zeta_{\rm cg}$ is in turn fixed by requiring that the compaction function reaches its critical value $C_{\rm c}\simeq0.5$, that is $C_1(r_m)\simeq0.67$ for type I perturbations. Using Eq.\,(\ref{eq:NG}), which gives $\zeta'=\zeta'_{\rm g}/(1-\zeta_{\rm g}/\mu)$, together with the profile of $C_{1{\rm b}}$, this reads
\begin{equation}\label{eq:thresh}
\frac{4}{3}\,\frac{\left|{\rm sinc}(k_{\rm max}r_m)-1\right|}{\ln(k_{\rm max}/k_{\rm min})}\,\frac{\zeta_{\rm cg}}{1-\zeta_{\rm cg}/\mu}\simeq0.67\,.
\end{equation}
Notice that for a broad spectrum the compaction at $r_m$ is suppressed with respect to the amplitude of $\zeta$ by the factor $1/\ln(k_{\rm max}/k_{\rm min})$: the infrared modes contribute to the variance of $\zeta$ but not to $C_1\propto r\zeta'$, which is sensitive only to the modes with $kr_m\gtrsim1$. As a consequence the threshold is considerably larger than for a peaked spectrum. Solving Eq.\,(\ref{eq:thresh}) at $\Delta=8.8$ for the range $1.5\lesssim\mu\lesssim5$ relevant for USR models one finds $\zeta_{\rm cg}\simeq1.3$--$3.2$ and therefore, from Eq.\,(\ref{eq:Ab}), $A_{\rm b}\simeq(1$--$6)\cdot10^{-3}$, whereas curvaton-like models with $\mu\simeq0.1$ would give $A_{\rm b}\sim10^{-6}$. As a benchmark value in the plot we use $A_{\rm b}=5\cdot10^{-3}$, corresponding to $\mu\simeq4$. We stress that at fixed abundance the amplitude of the power spectrum and the non-Gaussian parameters are degenerate, and models with different non-Gaussianity reproduce the same $f_{\rm PBH}$ with amplitudes differing by orders of magnitude\,\cite{Ferrante:2022mui}: the value of $A_{\rm b}$ should therefore be regarded as a benchmark, while the shape of the signal, a flat plateau extending over $\Delta$ decades with an ultraviolet cutoff at $f\simeq80\,$Hz set by $k_{\rm max}$, is a robust prediction of the mechanism.
 
Assuming a signal-to-noise ratio equal to $10$ and an observation time of $1$~year, we assess the detectability of the SIGW signals at ET. We consider ET in its standard ``xylophone'' layout, in which each detector is made of a high-frequency (HF) interferometer and a low-frequency (LF) interferometer, and we consider the different geometries currently under study~\cite{Branchesi:2023mws}, namely either a single-site observatory with a triangular configuration (${\rm ET_{TR}}$) made of three nested detectors, each composed of a LF and a HF interferometer and $10$~km-long arms; or a network of two L-shaped instruments (again, each made of a LF and a HF interferometer) with $15$~km-long arms each.%
\footnote{We follow the terminology of the 2020 ET Design Report Update, \url{https://apps.et-gw.eu/tds/?r=18715}: the high-frequency (HF) and low-frequency (LF) interferometers of the xylophone configuration are indeed referred to as ``interferometers''. The combination of a HF interferometer and a LF interferometer (whether in an L-shaped geometry, or with arms at $60^{\circ}$ as in the triangle configuration) is called a ``detector''. The whole ensemble of detectors is called an ``observatory''. So, ET in the triangle configuration is made of three detectors for a total of six interferometers, while in the 2L configuration it is made of two detectors, for a total of four interferometers.}
Currently, there are three candidate locations, which could in principle host either a triangle or an L-shaped detector in a 2L configuration: the Sos Enattos site in Sardinia; the Euroregion Meuse-Rhine (EMR) across Belgium, Germany and the Netherlands; and the Lusatia region in Saxony (Germany). For definiteness, we place the triangle configuration in Sardinia, and the 2L in Sardinia and in the EMR region.%
\footnote{Note, however, that, for a 2L configuration, placing the second detector in the candidate Saxony region would not affect the results, as the distance between the EMR and Saxony sites relative to the Sardinian location is quite similar.}
 
As the sensitivity to a stochastic background strongly depends on the relative orientation between detector pairs, we further consider scenarios in which the two L-shaped ET detectors are either aligned (${\rm ET_{LLa}}$) or misaligned (${\rm ET_{LLm}}$). As in Ref.~\cite{Branchesi:2023mws}, to take into account the curvature of the Earth, the relative orientation between the two detectors is defined with reference to the great circle that connects the two detectors. Denoting by $\beta$ the angle describing the relative orientation of the two detectors, defined with reference to this great circle, $\beta=0^{\circ}$ corresponds to the case where the arms of the two interferometers make the same angle with respect to the great circle, while $\beta=45^{\circ}$ corresponds to the case where one of the two interferometers is rotated by $45^{\circ}$ from the $\beta=0^{\circ}$ orientation. For $\beta=0^{\circ}$ (which we refer to as the ``2L-aligned'' configuration, ${\rm ET_{LLa}}$) the sensitivity to stochastic backgrounds is maximized, while for $\beta=45^{\circ}$ it is minimized, to the extent that the 2L configuration becomes completely blind to stochastic backgrounds in the limit $fd\to0$, where $f$ is the GW frequency and $d$ the distance between the two detectors. This is the opposite of what happens for compact binary coalescences (CBCs), where the accuracy of parameter reconstruction is maximized for $\beta=45^{\circ}$.
 
However, as discussed in Section~2 of Ref.~\cite{Branchesi:2023mws}, already a small misalignment angle of order a few degrees allows us to recover an interesting sensitivity to stochastic backgrounds, without basically affecting CBC reconstruction. In the following we use the same small misalignment angle as in Ref.~\cite{Branchesi:2023mws}, of about $2.5^{\circ}$, and we refer to this configuration as ``2L-misaligned'', ${\rm ET_{LLm}}$. Results for larger values of this relative angle will be intermediate between the results shown for the aligned and misaligned configurations.
 
We therefore consider a total of three configurations (triangle, 2L-aligned and 2L-misaligned), for which the power-law-integrated sensitivity curves, as defined in Ref.~\cite{Thrane:2013oya} (see also App.~A of Ref.~\cite{Branchesi:2023mws} for conventions and definitions), are computed using the public code \texttt{GWFAST}~\cite{Iacovelli:2022mbg}. The results are shown in Fig.~\ref{fig:situation}.
 
The corresponding SIGW spectrum lies in the sensitivity band of ET and at the same time may be tested simultaneously by future observatories such as LISA~\cite{LISA:2022kgy}, AION-km~\cite{Badurina:2019hst,Badurina:2021rgt,Abdalla:2024sst} and by the eLO experiment\,\cite{Foster:2025nzf}.

We wish to stress which feature of the signal is characteristic of the scenario. It is not the overall amplitude, which as discussed above is degenerate with the parameters of the non-Gaussianity at fixed abundance, but the shape: a plateau which is flat over $\Delta$ decades and is cut off in the ultraviolet at $f\simeq80\,$Hz, the frequency set by $k_{\rm max}$ and therefore by the mass of the light PBHs. A nearly flat stochastic background across the ET band is at present regarded as a signature of cosmic strings; the point of this work is that, if clusterogenesis takes place, such a signal is also produced by the PBHs which constitute the dark matter.
 
\section{Conclusions}
 
Giving an identity card to the DM is one of the main tasks in modern astroparticle physics. One attractive candidate is represented by PBHs, since this would not require physics beyond the Standard Model in terms of particle content. In this paper we have made the observation that the DM might be due entirely to asteroid-like PBHs in the mass range~(\ref{range}), originated by the collapse of much lighter PBHs; in fact so light that they would have evaporated otherwise. During the formation of such light PBHs, an ineluctable SIGW stochastic background is generated with frequencies falling not only in the LISA range, but also in the ET range, thus opening the possibility of revealing the presence of DM, albeit indirectly, by means of GW observations.
 
\section*{Acknowledgements}
 
We thank G.~Franciolini for participating at the early discussions of this work. A.J.I.\ research is funded by Tamkeen under the research grant to NYUAD ADHPG-AD457. A.R.\ and M.M.\ acknowledge support from the Swiss National Science Foundation (project number CRSII5\_213497). N.M.\ acknowledges support by the SwissMap National Center for Competence in Research. A.J.I.\ is grateful to the University of Geneva for the nice hospitality during the war in the Middle East.
 
\bibliography{main}

@article{Thrane:2013oya,
    author = "Thrane, Eric and Romano, Joseph D.",
    title = "{Sensitivity curves for searches for gravitational-wave backgrounds}",
    eprint = "1310.5300",
    archivePrefix = "arXiv",
    primaryClass = "astro-ph.IM",
    doi = "10.1103/PhysRevD.88.124032",
    journal = "Phys. Rev. D",
    volume = "88",
    number = "12",
    pages = "124032",
    year = "2013"
}

@article{Allegrini:2026jqt,
    author = {Allegrini, S. and Iovino, A. J. and Perna, G. and Veerm{\"a}e, H.},
    title = "{Memoirs of the curvaton: non-perturbative non-Gaussianity and supermassive primordial black holes}",
    eprint = "2606.28296",
    archivePrefix = "arXiv",
    primaryClass = "astro-ph.CO",
    month = "6",
    year = "2026"
}

@article{Auclair:2024jwj,
    author = "Auclair, Pierre and Blachier, Baptiste",
    title = "{Small-scale clustering of primordial black holes: Cloud-in-cloud and exclusion effects}",
    eprint = "2402.00600",
    archivePrefix = "arXiv",
    primaryClass = "astro-ph.CO",
    doi = "10.1103/PhysRevD.109.123538",
    journal = "Phys. Rev. D",
    volume = "109",
    number = "12",
    pages = "123538",
    year = "2024"
}

@article{LISACosmologyWorkingGroup:2025vdz,
    author = "Gammal, Jonas El and others",
    collaboration = "LISA Cosmology Working Group",
    title = "{Reconstructing primordial curvature perturbations via scalar-induced gravitational waves with LISA}",
    eprint = "2501.11320",
    archivePrefix = "arXiv",
    primaryClass = "astro-ph.CO",
    reportNumber = "CERN-TH-2024-217",
    doi = "10.1088/1475-7516/2025/05/062",
    journal = "JCAP",
    volume = "05",
    pages = "062",
    year = "2025"
}

@article{ET:2019dnz,
    author = "Maggiore, Michele and others",
    title = "{Science Case for the Einstein Telescope}",
    eprint = "1912.02622",
    archivePrefix = "arXiv",
    primaryClass = "astro-ph.CO",
    doi = "10.1088/1475-7516/2020/03/050",
    journal = "JCAP",
    volume = "03",
    pages = "050",
    year = "2020"
}

@article{ET:2025xjr,
    author = "Abac, Adrian and others",
    collaboration = "ET",
    title = "{The Science of the Einstein Telescope}",
    eprint = "2503.12263",
    archivePrefix = "arXiv",
    primaryClass = "gr-qc",
    reportNumber = "ET-0036C-25",
    month = "3",
    year = "2025"
}

@article{Punturo:2010zz,
    author = "Punturo, M. and others",
    editor = "Ricci, Fulvio",
    title = "{The Einstein Telescope: A third-generation gravitational wave observatory}",
    doi = "10.1088/0264-9381/27/19/194002",
    journal = "Class. Quant. Grav.",
    volume = "27",
    pages = "194002",
    year = "2010"
}

@article{Inui:2024fgk,
    author = "Inui, Ryoto and Joana, Cristian and Motohashi, Hayato and Pi, Shi and Tada, Yuichiro and Yokoyama, Shuichiro",
    title = "{Primordial black holes and induced gravitational waves from logarithmic non-Gaussianity}",
    eprint = "2411.07647",
    archivePrefix = "arXiv",
    primaryClass = "astro-ph.CO",
    doi = "10.1088/1475-7516/2025/03/021",
    journal = "JCAP",
    volume = "03",
    pages = "021",
    year = "2025"
}

@article{Tada:2015noa,
    author = "Tada, Yuichiro and Yokoyama, Shuichiro",
    title = "{Primordial black holes as biased tracers}",
    eprint = "1502.01124",
    archivePrefix = "arXiv",
    primaryClass = "astro-ph.CO",
    reportNumber = "IPMU-15-0014, RUP-15-3",
    doi = "10.1103/PhysRevD.91.123534",
    journal = "Phys. Rev. D",
    volume = "91",
    number = "12",
    pages = "123534",
    year = "2015"
}

@article{Delos:2024poq,
    author = "Delos, M. Sten and Rantala, Antti and Young, Sam and Schmidt, Fabian",
    title = "{Structure formation with primordial black holes: collisional dynamics, binaries, and gravitational waves}",
    eprint = "2410.01876",
    archivePrefix = "arXiv",
    primaryClass = "astro-ph.CO",
    doi = "10.1088/1475-7516/2024/12/005",
    journal = "JCAP",
    volume = "12",
    pages = "005",
    year = "2024"
}

@article{Bamber:2025gxj,
    author = "Bamber, Jamie and Shapiro, Stuart L. and Ruiz, Milton and Tsokaros, Antonios",
    title = "{Evolution of a black hole cluster in full general relativity}",
    eprint = "2505.01495",
    archivePrefix = "arXiv",
    primaryClass = "gr-qc",
    doi = "10.1103/8z1j-rzmx",
    journal = "Phys. Rev. D",
    volume = "112",
    number = "2",
    pages = "024046",
    year = "2025"
}

@article{Musco:2018rwt,
    author = "Musco, Ilia",
    title = "{Threshold for primordial black holes: Dependence on the shape of the cosmological perturbations}",
    eprint = "1809.02127",
    archivePrefix = "arXiv",
    primaryClass = "gr-qc",
    doi = "10.1103/PhysRevD.100.123524",
    journal = "Phys. Rev. D",
    volume = "100",
    number = "12",
    pages = "123524",
    year = "2019"
}

@article{Belotsky:2018wph,
    author = "Belotsky, Konstantin M. and Dokuchaev, Vyacheslav I. and Eroshenko, Yury N. and Esipova, Ekaterina A. and Khlopov, Maxim Yu. and Khromykh, Leonid A. and Kirillov, Alexander A. and Nikulin, Valeriy V. and Rubin, Sergey G. and Svadkovsky, Igor V.",
    title = "{Clusters of primordial black holes}",
    eprint = "1807.06590",
    archivePrefix = "arXiv",
    primaryClass = "astro-ph.CO",
    doi = "10.1140/epjc/s10052-019-6741-4",
    journal = "Eur. Phys. J. C",
    volume = "79",
    number = "3",
    pages = "246",
    year = "2019"
}

@article{Suyama:2019cst,
    author = "Suyama, Teruaki and Yokoyama, Shuichiro",
    title = "{Clustering of primordial black holes with non-Gaussian initial fluctuations}",
    eprint = "1906.04958",
    archivePrefix = "arXiv",
    primaryClass = "astro-ph.CO",
    doi = "10.1093/ptep/ptz105",
    journal = "PTEP",
    volume = "2019",
    number = "10",
    pages = "103E02",
    year = "2019"
}

@article{Matsubara:2019qzv,
    author = "Matsubara, Takahiko and Terada, Takahiro and Kohri, Kazunori and Yokoyama, Shuichiro",
    title = "{Clustering of primordial black holes formed in a matter-dominated epoch}",
    eprint = "1909.04053",
    archivePrefix = "arXiv",
    primaryClass = "astro-ph.CO",
    reportNumber = "KEK-TH-2155; KEK-Cosmo-243; IPMU19-0128, KEK-TH-2154; KEK-Cosmo-243; IPMU19-0128, KEK-TH-2154; KEK-Cosmo-243",
    doi = "10.1103/PhysRevD.100.123544",
    journal = "Phys. Rev. D",
    volume = "100",
    number = "12",
    pages = "123544",
    year = "2019"
}

@article{DeLuca:2020jug,
    author = "De Luca, V. and Desjacques, V. and Franciolini, G. and Riotto, A.",
    title = "{The clustering evolution of primordial black holes}",
    eprint = "2009.04731",
    archivePrefix = "arXiv",
    primaryClass = "astro-ph.CO",
    doi = "10.1088/1475-7516/2020/11/028",
    journal = "JCAP",
    volume = "11",
    pages = "028",
    year = "2020"
}

@article{DeLuca:2022bjs,
    author = "De Luca, Valerio and Franciolini, Gabriele and Riotto, Antonio",
    title = "{Heavy Primordial Black Holes from Strongly Clustered Light Black Holes}",
    eprint = "2210.14171",
    archivePrefix = "arXiv",
    primaryClass = "astro-ph.CO",
    doi = "10.1103/PhysRevLett.130.171401",
    journal = "Phys. Rev. Lett.",
    volume = "130",
    number = "17",
    pages = "171401",
    year = "2023"
}

@article{Crescimbeni:2025ywm,
    author = {Crescimbeni, F. and Desjacques, V. and Franciolini, G. and Ianniccari, A. and Iovino, A. J. and Perna, G. and Perrone, D. and Riotto, A. and Veerm{\"a}e, H.},
    title = "{The irrelevance of primordial black hole clustering in the LVK mass range}",
    eprint = "2502.01617",
    archivePrefix = "arXiv",
    primaryClass = "astro-ph.CO",
    reportNumber = "CERN-TH-2025-027",
    doi = "10.1088/1475-7516/2025/05/001",
    journal = "JCAP",
    volume = "05",
    pages = "001",
    year = "2025"
}

@article{Kolb:1993zz,
    author = "Kolb, Edward W. and Tkachev, Igor I.",
    title = "{Axion miniclusters and Bose stars}",
    eprint = "hep-ph/9303313",
    archivePrefix = "arXiv",
    reportNumber = "FERMILAB-PUB-93-066-A",
    doi = "10.1103/PhysRevLett.71.3051",
    journal = "Phys. Rev. Lett.",
    volume = "71",
    pages = "3051--3054",
    year = "1993"
}

@book{Misner:1973prb,
    author = "Misner, Charles W. and Thorne, K. S. and Wheeler, J. A.",
    title = "{Gravitation}",
    isbn = "978-0-7167-0344-0, 978-0-691-17779-3",
    publisher = "W. H. Freeman",
    address = "San Francisco",
    year = "1973"
}

@article{DeLuca:2022uvz,
    author = {De Luca, V. and Franciolini, G. and Riotto, A. and Veerm\"ae, H.},
    title = "{Ruling out Initial Primordial Black Hole Clustering}",
    eprint = "2208.01683",
    archivePrefix = "arXiv",
    primaryClass = "astro-ph.CO",
    journal = "Phys. Rev. Lett.",
    month = "8",
    year = "2022"
}

@article{Domenech:2021ztg,
    author = "Dom\`enech, Guillem",
    title = "{Scalar Induced Gravitational Waves Review}",
    eprint = "2109.01398",
    archivePrefix = "arXiv",
    primaryClass = "gr-qc",
    doi = "10.3390/universe7110398",
    journal = "Universe",
    volume = "7",
    number = "11",
    pages = "398",
    year = "2021"
}

@article{Inman:2019wvr,
    author = {Inman, Derek and Ali-Ha\"\i{}moud, Yacine},
    title = "{Early structure formation in primordial black hole cosmologies}",
    eprint = "1907.08129",
    archivePrefix = "arXiv",
    primaryClass = "astro-ph.CO",
    doi = "10.1103/PhysRevD.100.083528",
    journal = "Phys. Rev. D",
    volume = "100",
    number = "8",
    pages = "083528",
    year = "2019"
}

@article{Kolb:1994fi,
    author = "Kolb, Edward W. and Tkachev, Igor I.",
    title = "{Large amplitude isothermal fluctuations and high density dark matter clumps}",
    eprint = "astro-ph/9403011",
    archivePrefix = "arXiv",
    reportNumber = "FERMILAB-PUB-94-055-A",
    doi = "10.1103/PhysRevD.50.769",
    journal = "Phys. Rev. D",
    volume = "50",
    pages = "769--773",
    year = "1994"
}

@article{Green:2020jor,
    author = "Green, Anne M. and Kavanagh, Bradley J.",
    title = "{Primordial Black Holes as a dark matter candidate}",
    eprint = "2007.10722",
    archivePrefix = "arXiv",
    primaryClass = "astro-ph.CO",
    doi = "10.1088/1361-6471/abc534",
    journal = "J. Phys. G",
    volume = "48",
    number = "4",
    pages = "043001",
    year = "2021"
}

@article{Ali-Haimoud:2018dau,
    author = {Ali-Ha\"\i{}moud, Yacine},
    title = "{Correlation Function of High-Threshold Regions and Application to the Initial Small-Scale Clustering of Primordial Black Holes}",
    eprint = "1805.05912",
    archivePrefix = "arXiv",
    primaryClass = "astro-ph.CO",
    doi = "10.1103/PhysRevLett.121.081304",
    journal = "Phys. Rev. Lett.",
    volume = "121",
    number = "8",
    pages = "081304",
    year = "2018"
}

@article{Ballesteros:2018swv,
    author = "Ballesteros, Guillermo and Serpico, Pasquale D. and Taoso, Marco",
    title = "{On the merger rate of primordial black holes: effects of nearest neighbours distribution and clustering}",
    eprint = "1807.02084",
    archivePrefix = "arXiv",
    primaryClass = "astro-ph.CO",
    doi = "10.1088/1475-7516/2018/10/043",
    journal = "JCAP",
    volume = "10",
    pages = "043",
    year = "2018"
}

@article{Atal:2020igj,
    author = "Atal, Vicente and Sanglas, Albert and Triantafyllou, Nikolaos",
    title = "{LIGO/Virgo black holes and dark matter: The effect of spatial clustering}",
    eprint = "2007.07212",
    archivePrefix = "arXiv",
    primaryClass = "astro-ph.CO",
    doi = "10.1088/1475-7516/2020/11/036",
    journal = "JCAP",
    volume = "11",
    pages = "036",
    year = "2020"
}

@article{Desjacques:2018wuu,
    author = "Desjacques, Vincent and Riotto, Antonio",
    title = "{Spatial clustering of primordial black holes}",
    eprint = "1806.10414",
    archivePrefix = "arXiv",
    primaryClass = "astro-ph.CO",
    doi = "10.1103/PhysRevD.98.123533",
    journal = "Phys. Rev. D",
    volume = "98",
    number = "12",
    pages = "123533",
    year = "2018"
}

@article{DeLuca:2021hcf,
    author = "De Luca, V. and Franciolini, G. and Riotto, A.",
    title = "{Constraining the initial primordial black hole clustering with CMB distortion}",
    eprint = "2103.16369",
    archivePrefix = "arXiv",
    primaryClass = "astro-ph.CO",
    doi = "10.1103/PhysRevD.104.063526",
    journal = "Phys. Rev. D",
    volume = "104",
    number = "6",
    pages = "063526",
    year = "2021"
}

@article{Gorton:2022fyb,
    author = "Gorton, Matthew and Green, Anne M.",
    title = "{Effect of clustering on primordial black hole microlensing constraints}",
    eprint = "2203.04209",
    archivePrefix = "arXiv",
    primaryClass = "astro-ph.CO",
    month = "3",
    year = "2022"
}

@article{Petac:2022rio,
    author = "Peta\v{c}, Mihael and Lavalle, Julien and Jedamzik, Karsten",
    title = "{Microlensing constraints on clustered primordial black holes}",
    eprint = "2201.02521",
    archivePrefix = "arXiv",
    primaryClass = "astro-ph.CO",
    doi = "10.1103/PhysRevD.105.083520",
    journal = "Phys. Rev. D",
    volume = "105",
    number = "8",
    pages = "083520",
    year = "2022"
}

@book{1987gady.book.....B,
       author = {{Binney}, James and {Tremaine}, Scott},
        title = "{Galactic dynamics}",
         year = 1987,
       adsurl = {https://ui.adsabs.harvard.edu/abs/1987gady.book.....B},
    publisher = {Princeton University Press},
}

@article{Young:2019gfc,
    author = "Young, Sam and Byrnes, Christian T.",
    title = "{Initial clustering and the primordial black hole merger rate}",
    eprint = "1910.06077",
    archivePrefix = "arXiv",
    primaryClass = "astro-ph.CO",
    doi = "10.1088/1475-7516/2020/03/004",
    journal = "JCAP",
    volume = "03",
    pages = "004",
    year = "2020"
}

@article{Carr:2026hot,
    author = {Carr, Bernard and Iovino, Antonio J. and Perna, Gabriele and Vaskonen, Ville and Veerm{\"a}e, Hardi},
    title = "{Primordial black holes: constraints, potential evidence and prospects}",
    eprint = "2601.06024",
    archivePrefix = "arXiv",
    primaryClass = "astro-ph.CO",
    doi = "10.1007/s40766-026-00080-z",
    journal = "Riv. Nuovo Cim.",
    volume = "49",
    number = "5",
    pages = "225--274",
    year = "2026"
}

@article{Animali:2024jiz,
    author = "Animali, Chiara and Vennin, Vincent",
    title = "{Clustering of primordial black holes from quantum diffusion during inflation}",
    eprint = "2402.08642",
    archivePrefix = "arXiv",
    primaryClass = "astro-ph.CO",
    doi = "10.1088/1475-7516/2024/08/026",
    journal = "JCAP",
    volume = "08",
    pages = "026",
    year = "2024"
}

@article{Arbey:2019jmj,
    author = "Arbey, Alexandre and Auffinger, J{\'e}r{\'e}my and Silk, Joseph",
    title = "{Evolution of primordial black hole spin due to Hawking radiation}",
    eprint = "1906.04196",
    archivePrefix = "arXiv",
    primaryClass = "astro-ph.CO",
    reportNumber = "CERN-TH-2019-068",
    doi = "10.1093/mnras/staa765",
    journal = "Mon. Not. Roy. Astron. Soc.",
    volume = "494",
    number = "1",
    pages = "1257--1262",
    year = "2020"
}

@article{Iovino:2025xkq,
    author = "Iovino, Antonio Junior and Perna, Gabriele and Perrone, Davide and Racco, Davide and Riotto, Antonio",
    title = "{Understanding the nature of scalar-induced gravitational waves}",
    eprint = "2509.24774",
    archivePrefix = "arXiv",
    primaryClass = "gr-qc",
    doi = "10.1088/1475-7516/2026/03/001",
    journal = "JCAP",
    volume = "03",
    pages = "001",
    year = "2026"
}

@article{Escriva:2022duf,
    author = "Escriv\`a, Albert and Kuhnel, Florian and Tada, Yuichiro",
    title = "{Primordial Black Holes}",
    eprint = "2211.05767",
    archivePrefix = "arXiv",
    primaryClass = "astro-ph.CO",
    doi = "10.1016/B978-0-32-395636-9.00012-8",
    month = "11",
    year = "2022"
}

@article{Iovino:2024tyg,
    author = {Iovino, A. J. and Perna, G. and Riotto, A. and Veerm\"ae, H.},
    title = "{Curbing PBHs with PTAs}",
    eprint = "2406.20089",
    archivePrefix = "arXiv",
    primaryClass = "astro-ph.CO",
    doi = "10.1088/1475-7516/2024/10/050",
    journal = "JCAP",
    volume = "10",
    pages = "050",
    year = "2024"
}

@article{Enqvist:2005pg,
    author = "Enqvist, Kari and Nurmi, Sami",
    title = "{Non-gaussianity in curvaton models with nearly quadratic potential}",
    eprint = "astro-ph/0508573",
    archivePrefix = "arXiv",
    reportNumber = "HIP-2005-33-TH",
    doi = "10.1088/1475-7516/2005/10/013",
    journal = "JCAP",
    volume = "10",
    pages = "013",
    year = "2005"
}

@article{Lyth:2006gd,
    author = "Lyth, David H.",
    title = "{Non-gaussianity and cosmic uncertainty in curvaton-type models}",
    eprint = "astro-ph/0602285",
    archivePrefix = "arXiv",
    doi = "10.1088/1475-7516/2006/06/015",
    journal = "JCAP",
    volume = "06",
    pages = "015",
    year = "2006"
}

@article{Cai:2018dkf,
    author = "Cai, Yi-Fu and Chen, Xingang and Namjoo, Mohammad Hossein and Sasaki, Misao and Wang, Dong-Gang and Wang, Ziwei",
    title = "{Revisiting non-Gaussianity from non-attractor inflation models}",
    eprint = "1712.09998",
    archivePrefix = "arXiv",
    primaryClass = "astro-ph.CO",
    reportNumber = "MIT-CTP-4974, YITP-17-133",
    doi = "10.1088/1475-7516/2018/05/012",
    journal = "JCAP",
    volume = "05",
    pages = "012",
    year = "2018"
}

@article{Cai:2018dig,
	title        = {{Gravitational Waves Induced by non-Gaussian Scalar Perturbations}},
	author       = {Cai, Rong-gen and Pi, Shi and Sasaki, Misao},
	year         = 2019,
	journal      = {Phys. Rev. Lett.},
	volume       = 122,
	number       = 20,
	pages        = 201101,
	doi          = {10.1103/PhysRevLett.122.201101},
	eprint       = {1810.11000},
	archiveprefix = {arXiv},
	primaryclass = {astro-ph.CO},
	reportnumber = {IPMU18-0172, YITP-18-114}
}

@article{Cai:2019elf,
	title        = {{Pulsar Timing Array Constraints on the Induced Gravitational Waves}},
	author       = {Cai, Rong-Gen and Pi, Shi and Wang, Shao-Jiang and Yang, Xing-Yu},
	year         = 2019,
	journal      = {JCAP},
	volume       = 10,
	pages        = {059},
	doi          = {10.1088/1475-7516/2019/10/059},
	eprint       = {1907.06372},
	archiveprefix = {arXiv},
	primaryclass = {astro-ph.CO}
}

@article{Hajkarim:2019nbx,
	title        = {{Thermal History of the Early Universe and Primordial Gravitational Waves from Induced Scalar Perturbations}},
	author       = {Hajkarim, Fazlollah and Schaffner-Bielich, J\"urgen},
	year         = 2020,
	journal      = {Phys. Rev. D},
	volume       = 101,
	number       = 4,
	pages        = {043522},
	doi          = {10.1103/PhysRevD.101.043522},
	eprint       = {1910.12357},
	archiveprefix = {arXiv},
	primaryclass = {hep-ph}
}

@article{Atal:2021jyo,
	title        = {{Probing non-Gaussianities with the high frequency tail of induced gravitational waves}},
	author       = {Atal, Vicente and Dom\`enech, Guillem},
	year         = 2021,
	journal      = {JCAP},
	volume       = {06},
	pages        = {001},
	doi          = {10.1088/1475-7516/2021/06/001},
	note         = {[Erratum: JCAP 10, E01 (2023)]},
	eprint       = {2103.01056},
	archiveprefix = {arXiv},
	primaryclass = {astro-ph.CO}
}

@article{Yuan:2020iwf,
	title        = {{Gravitational waves induced by the local-type non-Gaussian curvature perturbations}},
	author       = {Yuan, Chen and Huang, Qing-Guo},
	year         = 2021,
	journal      = {Phys. Lett. B},
	volume       = 821,
	pages        = 136606,
	doi          = {10.1016/j.physletb.2021.136606},
	eprint       = {2007.10686},
	archiveprefix = {arXiv},
	primaryclass = {astro-ph.CO}
}

@article{Domenech:2021and,
	title        = {{Gravitational waves from dark matter isocurvature}},
	author       = {Dom\`enech, Guillem and Passaglia, Samuel and Renaux-Petel, S\'ebastien},
	year         = 2022,
	journal      = {JCAP},
	volume       = {03},
	number       = {03},
	pages        = {023},
	doi          = {10.1088/1475-7516/2022/03/023},
	eprint       = {2112.10163},
	archiveprefix = {arXiv},
	primaryclass = {astro-ph.CO},
	reportnumber = {ET-0466A-21}
}

@article{Garcia-Saenz:2022tzu,
	title        = {{No-go theorem for scalar-trispectrum-induced gravitational waves}},
	author       = {Garcia-Saenz, Sebastian and Pinol, Lucas and Renaux-Petel, S\'ebastien and Werth, Denis},
	year         = 2023,
	journal      = {JCAP},
	volume       = {03},
	pages        = {057},
	doi          = {10.1088/1475-7516/2023/03/057},
	eprint       = {2207.14267},
	archiveprefix = {arXiv},
	primaryclass = {astro-ph.CO}
}

@article{Liu:2023ymk,
	title        = {{Implications for the non-Gaussianity of curvature perturbation from pulsar timing arrays}},
	author       = {Liu, Lang and Chen, Zu-Cheng and Huang, Qing-Guo},
	year         = 2024,
	journal      = {Phys. Rev. D},
	volume       = 109,
	number       = 6,
	pages        = {L061301},
	doi          = {10.1103/PhysRevD.109.L061301},
	eprint       = {2307.01102},
	archiveprefix = {arXiv},
	primaryclass = {astro-ph.CO}
}

@article{Yuan:2023ofl,
	title        = {{Full analysis of the scalar-induced gravitational waves for the curvature perturbation with local-type non-Gaussianities}},
	author       = {Yuan, Chen and Meng, De-Shuang and Huang, Qing-Guo},
	year         = 2023,
	journal      = {JCAP},
	volume       = 12,
	pages        = {036},
	doi          = {10.1088/1475-7516/2023/12/036},
	eprint       = {2308.07155},
	archiveprefix = {arXiv},
	primaryclass = {astro-ph.CO}
}

@article{Li:2023xtl,
	title        = {{Complete analysis of the background and anisotropies of scalar-induced gravitational waves: primordial non-Gaussianity f $_{NL}$ and g $_{NL}$ considered}},
	author       = {Li, Jun-Peng and Wang, Sai and Zhao, Zhi-Chao and Kohri, Kazunori},
	year         = 2024,
	journal      = {JCAP},
	volume       = {06},
	pages        = {039},
	doi          = {10.1088/1475-7516/2024/06/039},
	eprint       = {2309.07792},
	archiveprefix = {arXiv},
	primaryclass = {astro-ph.CO},
	reportnumber = {KEK-Cosmo-0326, KEK-TH-2556, KEK-QUP-2023-0024}
}

@article{Unal:2018yaa,
	title        = {{Imprints of Primordial Non-Gaussianity on Gravitational Wave Spectrum}},
	author       = {Unal, Caner},
	year         = 2019,
	journal      = {Phys. Rev. D},
	volume       = 99,
	number       = 4,
	pages        = {041301},
	doi          = {10.1103/PhysRevD.99.041301},
	eprint       = {1811.09151},
	archiveprefix = {arXiv},
	primaryclass = {astro-ph.CO}
}

@article{Adshead:2021hnm,
	title        = {{Non-Gaussianity and the induced gravitational wave background}},
	author       = {Adshead, Peter and Lozanov, Kaloian D. and Weiner, Zachary J.},
	year         = 2021,
	journal      = {JCAP},
	volume       = 10,
	pages        = {080},
	doi          = {10.1088/1475-7516/2021/10/080},
	eprint       = {2105.01659},
	archiveprefix = {arXiv},
	primaryclass = {astro-ph.CO}
}

@article{Perna:2024ehx,
	title        = {{Fully non-Gaussian Scalar-Induced Gravitational Waves}},
	author       = {Perna, Gabriele and Testini, Chiara and Ricciardone, Angelo and Matarrese, Sabino},
	year         = 2024,
	journal      = {JCAP},
	volume       = {05},
	pages        = {086},
	doi          = {10.1088/1475-7516/2024/05/086},
	eprint       = {2403.06962},
	archiveprefix = {arXiv},
	primaryclass = {astro-ph.CO}
}

@article{LISACosmologyWorkingGroup:2022jok,
	title        = {{Cosmology with the Laser Interferometer Space Antenna}},
	author       = {Auclair, Pierre and others},
	year         = 2023,
	journal      = {Living Rev. Rel.},
	volume       = 26,
	number       = 1,
	pages        = 5,
	doi          = {10.1007/s41114-023-00045-2},
	collaboration = {LISA Cosmology Working Group},
	eprint       = {2204.05434},
	archiveprefix = {arXiv},
	primaryclass = {astro-ph.CO},
	reportnumber = {LISA CosWG-22-03, FERMILAB-PUB-22-349-SCD}
}

@article{Tomita:1975kj,
	title        = {{Evolution of Irregularities in a Chaotic Early Universe}},
	author       = {Tomita, Kenji},
	year         = 1975,
	journal      = {Prog. Theor. Phys.},
	volume       = 54,
	pages        = 730,
	doi          = {10.1143/PTP.54.730},
	reportnumber = {RRK 75-3}
}

@article{Matarrese:1993zf,
	title        = {{General relativistic dynamics of irrotational dust: Cosmological implications}},
	author       = {Matarrese, Sabino and Pantano, Ornella and Saez, Diego},
	year         = 1994,
	journal      = {Phys. Rev. Lett.},
	volume       = 72,
	pages        = {320--323},
	doi          = {10.1103/PhysRevLett.72.320},
	eprint       = {astro-ph/9310036},
	archiveprefix = {arXiv},
	reportnumber = {DFPD-93-A-67}
}

@article{Acquaviva:2002ud,
	title        = {{Second order cosmological perturbations from inflation}},
	author       = {Acquaviva, Viviana and Bartolo, Nicola and Matarrese, Sabino and Riotto, Antonio},
	year         = 2003,
	journal      = {Nucl. Phys. B},
	volume       = 667,
	pages        = {119--148},
	doi          = {10.1016/S0550-3213(03)00550-9},
	eprint       = {astro-ph/0209156},
	archiveprefix = {arXiv},
	reportnumber = {DFPD-A-02-21}
}

@article{Mollerach:2003nq,
	title        = {{CMB polarization from secondary vector and tensor modes}},
	author       = {Mollerach, Silvia and Harari, Diego and Matarrese, Sabino},
	year         = 2004,
	journal      = {Phys. Rev. D},
	volume       = 69,
	pages        = {063002},
	doi          = {10.1103/PhysRevD.69.063002},
	eprint       = {astro-ph/0310711},
	archiveprefix = {arXiv}
}

@article{Ananda:2006af,
	title        = {{The Cosmological gravitational wave background from primordial density perturbations}},
	author       = {Ananda, Kishore N. and Clarkson, Chris and Wands, David},
	year         = 2007,
	journal      = {Phys. Rev. D},
	volume       = 75,
	pages        = 123518,
	doi          = {10.1103/PhysRevD.75.123518},
	eprint       = {gr-qc/0612013},
	archiveprefix = {arXiv}
}

@article{Baumann:2007zm,
	title        = {{Gravitational Wave Spectrum Induced by Primordial Scalar Perturbations}},
	author       = {Baumann, Daniel and Steinhardt, Paul J. and Takahashi, Keitaro and Ichiki, Kiyotomo},
	year         = 2007,
	journal      = {Phys. Rev. D},
	volume       = 76,
	pages        = {084019},
	doi          = {10.1103/PhysRevD.76.084019},
	eprint       = {hep-th/0703290},
	archiveprefix = {arXiv}
}

@article{Abdalla:2024sst,
    author = "Abdalla, Adam and others",
    title = "{Terrestrial Very-Long-Baseline Atom Interferometry: summary of the second workshop}",
    eprint = "2412.14960",
    archivePrefix = "arXiv",
    primaryClass = "hep-ex",
    doi = "10.1140/epjqt/s40507-025-00344-3",
    journal = "EPJ Quant. Technol.",
    volume = "12",
    number = "1",
    pages = "42",
    year = "2025"
}

@article{Foster:2025nzf,
    author = "Foster, Joshua W. and Blas, Diego and Bourgoin, Adrien and Hees, Aurelien and Herrero-Valea, M{\'\i}riam and Jenkins, Alexander C. and Xue, Xiao",
    title = "{Discovering $\mu$Hz gravitational waves and ultra-light dark matter with binary resonances}",
    eprint = "2504.15334",
    archivePrefix = "arXiv",
    primaryClass = "astro-ph.CO",
    reportNumber = "FERMILAB-PUB-25-0091-T",
    month = "4",
    year = "2025"
}

@article{Carr:1974nx,
	title        = {{Black holes in the early Universe}},
	author       = {Carr, Bernard J. and Hawking, S. W.},
	year         = 1974,
	journal      = {Mon. Not. Roy. Astron. Soc.},
	volume       = 168,
	pages        = {399--415},
	doi          = {10.1093/mnras/168.2.399}
}

@article{Carr:1975qj,
	title        = {{The Primordial black hole mass spectrum}},
	author       = {Carr, Bernard J.},
	year         = 1975,
	journal      = {Astrophys. J.},
	volume       = 201,
	pages        = {1--19},
	doi          = {10.1086/153853}
}

@article{Franciolini:2022pav,
    author = "Franciolini, Gabriele and Urbano, Alfredo",
    title = "{Primordial black hole dark matter from inflation: The reverse engineering approach}",
    eprint = "2207.10056",
    archivePrefix = "arXiv",
    primaryClass = "astro-ph.CO",
    doi = "10.1103/PhysRevD.106.123519",
    journal = "Phys. Rev. D",
    volume = "106",
    number = "12",
    pages = "123519",
    year = "2022"
}

@article{Atal:2019cdz,
	title        = {{Primordial black hole formation with non-Gaussian curvature perturbations}},
	author       = {Atal, Vicente and Garriga, Jaume and Marcos-Caballero, Airam},
	year         = 2019,
	journal      = {JCAP},
	volume       = {09},
	pages        = {073},
	doi          = {10.1088/1475-7516/2019/09/073},
	eprint       = {1905.13202},
	archiveprefix = {arXiv},
	primaryclass = {astro-ph.CO}
}

@article{Ando:2018nge,
	title        = {{Formation of primordial black holes in an axionlike curvaton model}},
	author       = {Ando, Kenta and Kawasaki, Masahiro and Nakatsuka, Hiromasa},
	year         = 2018,
	journal      = {Phys. Rev. D},
	volume       = 98,
	number       = 8,
	pages        = {083508},
	doi          = {10.1103/PhysRevD.98.083508},
	eprint       = {1805.07757},
	archiveprefix = {arXiv},
	primaryclass = {astro-ph.CO},
	reportnumber = {IPMU18-0087}
}

@article{Ferrante:2023bgz,
	title        = {{Primordial black holes in the curvaton model: possible connections to pulsar timing arrays and dark matter}},
	author       = {Ferrante, Giacomo and Franciolini, Gabriele and Iovino, Junior., Antonio and Urbano, Alfredo},
	year         = 2023,
	journal      = {JCAP},
	volume       = {06},
	pages        = {057},
	doi          = {10.1088/1475-7516/2023/06/057},
	eprint       = {2305.13382},
	archiveprefix = {arXiv},
	primaryclass = {astro-ph.CO}
}

@article{Young:2022phe,
    author = "Young, Sam",
    title = "{Peaks and primordial black holes: the~effect of non-Gaussianity}",
    eprint = "2201.13345",
    archivePrefix = "arXiv",
    primaryClass = "astro-ph.CO",
    doi = "10.1088/1475-7516/2022/05/037",
    journal = "JCAP",
    volume = "05",
    number = "05",
    pages = "037",
    year = "2022"
}

@article{Chisholm:2005vm,
    author = "Chisholm, James R.",
    title = "{Clustering of primordial black holes: basic results}",
    eprint = "astro-ph/0509141",
    archivePrefix = "arXiv",
    doi = "10.1103/PhysRevD.73.083504",
    journal = "Phys. Rev. D",
    volume = "73",
    pages = "083504",
    year = "2006"
}

@article{Young:2019yug,
    author = "Young, Sam and Musco, Ilia and Byrnes, Christian T.",
    title = "{Primordial black hole formation and abundance: contribution from the non-linear relation between the density and curvature perturbation}",
    eprint = "1904.00984",
    archivePrefix = "arXiv",
    primaryClass = "astro-ph.CO",
    doi = "10.1088/1475-7516/2019/11/012",
    journal = "JCAP",
    volume = "11",
    pages = "012",
    year = "2019"
}

@article{Espinosa:2018eve,
	title        = {{A Cosmological Signature of the SM Higgs Instability: Gravitational Waves}},
	author       = {Espinosa, Jos\'e Ram\'on and Racco, Davide and Riotto, Antonio},
	year         = 2018,
	journal      = {JCAP},
	volume       = {09},
	pages        = {012},
	doi          = {10.1088/1475-7516/2018/09/012},
	eprint       = {1804.07732},
	archiveprefix = {arXiv},
	primaryclass = {hep-ph}
}

@article{Karam:2022nym,
	title        = {{Anatomy of single-field inflationary models for primordial black holes}},
	author       = {Karam, Alexandros and Koivunen, Niko and Tomberg, Eemeli and Vaskonen, Ville and Veerm\"ae, Hardi},
	year         = 2023,
	journal      = {JCAP},
	volume       = {03},
	pages        = {013},
	doi          = {10.1088/1475-7516/2023/03/013},
	eprint       = {2205.13540},
	archiveprefix = {arXiv},
	primaryclass = {astro-ph.CO}
}

@article{Sasaki:2006kq,
	title        = {{Non-Gaussianity of the primordial perturbation in the curvaton model}},
	author       = {Sasaki, Misao and Valiviita, Jussi and Wands, David},
	year         = 2006,
	journal      = {Phys. Rev. D},
	volume       = 74,
	pages        = 103003,
	doi          = {10.1103/PhysRevD.74.103003},
	eprint       = {astro-ph/0607627},
	archiveprefix = {arXiv},
	reportnumber = {YITP-06-33}
}

@article{Atal:2018neu,
	title        = {{The role of non-gaussianities in Primordial Black Hole formation}},
	author       = {Atal, Vicente and Germani, Cristiano},
	year         = 2019,
	journal      = {Phys. Dark Univ.},
	volume       = 24,
	pages        = 100275,
	doi          = {10.1016/j.dark.2019.100275},
	eprint       = {1811.07857},
	archiveprefix = {arXiv},
	primaryclass = {astro-ph.CO},
	reportnumber = {ICCUB-18-022}
}

@article{Sasaki:2018dmp,
	title        = {{Primordial black holes\textemdash{}perspectives in gravitational wave astronomy}},
	author       = {Sasaki, Misao and Suyama, Teruaki and Tanaka, Takahiro and Yokoyama, Shuichiro},
	year         = 2018,
	journal      = {Class. Quant. Grav.},
	volume       = 35,
	number       = 6,
	pages        = {063001},
	doi          = {10.1088/1361-6382/aaa7b4},
	eprint       = {1801.05235},
	archiveprefix = {arXiv},
	primaryclass = {astro-ph.CO}
}

@article{Carr:2020gox,
	title        = {{Constraints on primordial black holes}},
	author       = {Carr, Bernard and Kohri, Kazunori and Sendouda, Yuuiti and Yokoyama, Jun'ichi},
	year         = 2021,
	journal      = {Rept. Prog. Phys.},
	volume       = 84,
	number       = 11,
	pages        = 116902,
	doi          = {10.1088/1361-6633/ac1e31},
	eprint       = {2002.12778},
	archiveprefix = {arXiv},
	primaryclass = {astro-ph.CO},
	reportnumber = {RESCEU-03/20; KEK-Cosmo-249; KEK-TH-2199; IPMU20-0024}
}

@article{Bartolo:2018evs,
	title        = {{Primordial Black Hole Dark Matter: LISA Serendipity}},
	author       = {Bartolo, N. and De Luca, V. and Franciolini, G. and Lewis, A. and Peloso, M. and Riotto, A.},
	year         = 2019,
	journal      = {Phys. Rev. Lett.},
	volume       = 122,
	number       = 21,
	pages        = 211301,
	doi          = {10.1103/PhysRevLett.122.211301},
	eprint       = {1810.12218},
	archiveprefix = {arXiv},
	primaryclass = {astro-ph.CO}
}

@article{MoradinezhadDizgah:2019wjf,
	title        = {{Primordial Black Holes from Broad Spectra: Abundance and Clustering}},
	author       = {Moradinezhad Dizgah, Azadeh and Franciolini, Gabriele and Riotto, Antonio},
	year         = 2019,
	journal      = {JCAP},
	volume       = 11,
	pages        = {001},
	doi          = {10.1088/1475-7516/2019/11/001},
	eprint       = {1906.08978},
	archiveprefix = {arXiv},
	primaryclass = {astro-ph.CO}
}

@article{Leach:2001zf,
	title        = {{Enhancement of superhorizon scale inflationary curvature perturbations}},
	author       = {Leach, Samuel M and Sasaki, Misao and Wands, David and Liddle, Andrew R},
	year         = 2001,
	journal      = {Phys. Rev. D},
	volume       = 64,
	pages        = {023512},
	doi          = {10.1103/PhysRevD.64.023512},
	eprint       = {astro-ph/0101406},
	archiveprefix = {arXiv}
}

@article{LISACosmologyWorkingGroup:2024hsc,
    author = "Braglia, Matteo and others",
    collaboration = "LISA Cosmology Working Group",
    title = "{Gravitational waves from inflation in LISA: reconstruction pipeline and physics interpretation}",
    eprint = "2407.04356",
    archivePrefix = "arXiv",
    primaryClass = "astro-ph.CO",
    reportNumber = "LISA-COSWG-24-03, CERN-TH-2024-072",
    doi = "10.1088/1475-7516/2024/11/032",
    journal = "JCAP",
    volume = "11",
    pages = "032",
    year = "2024"
}

@article{Iovino:2024sgs,
    author = "Iovino, A. J. and Matarrese, S. and Perna, G. and Ricciardone, A. and Riotto, A.",
    title = "{How well do we know the scalar-induced gravitational waves?}",
    eprint = "2412.06764",
    archivePrefix = "arXiv",
    primaryClass = "astro-ph.CO",
    doi = "10.1016/j.physletb.2025.140039",
    journal = "Phys. Lett. B",
    volume = "872",
    pages = "140039",
    year = "2026"
}

@article{Branchesi:2023mws,
    author = "Branchesi, Marica and others",
    title = "{Science with the Einstein Telescope: a comparison of different designs}",
    eprint = "2303.15923",
    archivePrefix = "arXiv",
    primaryClass = "gr-qc",
    reportNumber = "ET-0084A-23",
    doi = "10.1088/1475-7516/2023/07/068",
    journal = "JCAP",
    volume = "07",
    pages = "068",
    year = "2023"
}

@article{LISA:2022kgy,
    author = "Arun, K. G. and others",
    collaboration = "LISA",
    title = "{New horizons for fundamental physics with LISA}",
    eprint = "2205.01597",
    archivePrefix = "arXiv",
    primaryClass = "gr-qc",
    doi = "10.1007/s41114-022-00036-9",
    journal = "Living Rev. Rel.",
    volume = "25",
    number = "1",
    pages = "4",
    year = "2022"
}

@article{Iovino:2025cdy,
    author = {Iovino, Junior., Antonio and Perna, Gabriele and Veerm{\"a}e, Hardi},
    title = "{The impact of non-Gaussianity when searching for Primordial Black Holes with LISA}",
    eprint = "2512.13648",
    archivePrefix = "arXiv",
    primaryClass = "astro-ph.CO",
    doi = "10.1088/1475-7516/2026/05/027",
    journal = "JCAP",
    volume = "05",
    pages = "027",
    year = "2026"
}

@article{Blas:2026xws,
    author = "Blas, D. and Foster, J. W. and Gouttenoire, Y. and Iovino, A. J. and Musco, I. and Trifinopoulos, S. and Vanvlasselaer, M.",
    title = "{The Dark Side of the Moon: Listening to Scalar-Induced Gravitational Waves}",
    eprint = "2602.12252",
    archivePrefix = "arXiv",
    primaryClass = "astro-ph.CO",
    reportNumber = "CERN-TH-2026-021",
    month = "2",
    year = "2026"
}

@article{Bartolo:2018rku,
    author = "Bartolo, N. and De Luca, V. and Franciolini, G. and Peloso, M. and Racco, D. and Riotto, A.",
    title = "{Testing primordial black holes as dark matter with LISA}",
    eprint = "1810.12224",
    archivePrefix = "arXiv",
    primaryClass = "astro-ph.CO",
    doi = "10.1103/PhysRevD.99.103521",
    journal = "Phys. Rev. D",
    volume = "99",
    number = "10",
    pages = "103521",
    year = "2019"
}

@article{Lyth:2002my,
    author = "Lyth, David H. and Ungarelli, Carlo and Wands, David",
    title = "{The Primordial density perturbation in the curvaton scenario}",
    eprint = "astro-ph/0208055",
    archivePrefix = "arXiv",
    reportNumber = "PU-ICG-02-15",
    doi = "10.1103/PhysRevD.67.023503",
    journal = "Phys. Rev. D",
    volume = "67",
    pages = "023503",
    year = "2003"
}

@article{Ferrante:2022mui,
    author = "Ferrante, Giacomo and Franciolini, Gabriele and Iovino, Junior., Antonio and Urbano, Alfredo",
    title = "{Primordial non-Gaussianity up to all orders: Theoretical aspects and implications for primordial black hole models}",
    eprint = "2211.01728",
    archivePrefix = "arXiv",
    primaryClass = "astro-ph.CO",
    doi = "10.1103/PhysRevD.107.043520",
    journal = "Phys. Rev. D",
    volume = "107",
    number = "4",
    pages = "043520",
    year = "2023"
}

@article{Harada:2015yda,
    author = "Harada, Tomohiro and Yoo, Chul-Moon and Nakama, Tomohiro and Koga, Yasutaka",
    title = "{Cosmological long-wavelength solutions and primordial black hole formation}",
    eprint = "1503.03934",
    archivePrefix = "arXiv",
    primaryClass = "gr-qc",
    reportNumber = "RUP-15-5, RESCEU-4-15",
    doi = "10.1103/PhysRevD.91.084057",
    journal = "Phys. Rev. D",
    volume = "91",
    number = "8",
    pages = "084057",
    year = "2015"
}

@article{Badurina:2021rgt,
    author = "Badurina, Leonardo and Buchmueller, Oliver and Ellis, John and Lewicki, Marek and McCabe, Christopher and Vaskonen, Ville",
    title = "{Prospective sensitivities of atom interferometers to gravitational waves and ultralight
 dark matter}",
    eprint = "2108.02468",
    archivePrefix = "arXiv",
    primaryClass = "gr-qc",
    reportNumber = "AION-REPORT/2021-04, KCL-PH-TH/2021-61, CERN-TH-2021-116",
    doi = "10.1098/rsta.2021.0060",
    journal = "Phil. Trans. A. Math. Phys. Eng. Sci.",
    volume = "380",
    number = "2216",
    pages = "20210060",
    year = "2021"
}

@article{Badurina:2019hst,
    author = "Badurina, L. and others",
    title = "{AION: An Atom Interferometer Observatory and Network}",
    eprint = "1911.11755",
    archivePrefix = "arXiv",
    primaryClass = "astro-ph.CO",
    reportNumber = "AION-2019-001, CERN-TH-2019-199",
    doi = "10.1088/1475-7516/2020/05/011",
    journal = "JCAP",
    volume = "05",
    pages = "011",
    year = "2020"
}

@article{Shibata:1999zs,
    author = "Shibata, Masaru and Sasaki, Misao",
    title = "{Black hole formation in the Friedmann universe: Formulation and computation in numerical relativity}",
    eprint = "gr-qc/9905064",
    archivePrefix = "arXiv",
    reportNumber = "OU-TAP-93",
    doi = "10.1103/PhysRevD.60.084002",
    journal = "Phys. Rev. D",
    volume = "60",
    pages = "084002",
    year = "1999"
}

@article{Pi:2022ysn,
    author = "Pi, Shi and Sasaki, Misao",
    title = "{Logarithmic Duality of the Curvature Perturbation}",
    eprint = "2211.13932",
    archivePrefix = "arXiv",
    primaryClass = "astro-ph.CO",
    reportNumber = "IPMU22-0060, YITP-22-144",
    doi = "10.1103/PhysRevLett.131.011002",
    journal = "Phys. Rev. Lett.",
    volume = "131",
    number = "1",
    pages = "011002",
    year = "2023"
}

@article{Frosina:2023nxu,
    author = "Frosina, Lorenzo and Urbano, Alfredo",
    title = "{Inflationary interpretation of the nHz gravitational-wave background}",
    eprint = "2308.06915",
    archivePrefix = "arXiv",
    primaryClass = "astro-ph.CO",
    doi = "10.1103/PhysRevD.108.103544",
    journal = "Phys. Rev. D",
    volume = "108",
    number = "10",
    pages = "103544",
    year = "2023"
}

@ARTICLE{1986ApJ...304...15B,
       author = {{Bardeen}, J.~M. and {Bond}, J.~R. and {Kaiser}, N. and {Szalay}, A.~S.},
        title = "{The Statistics of Peaks of Gaussian Random Fields}",
      journal = {apj},
         year = 1986,
        month = may,
       volume = {304},
        pages = {15},
          doi = {10.1086/164143},
       adsurl = {https://ui.adsabs.harvard.edu/abs/1986ApJ...304...15B}
}

@article{DeLuca:2019qsy,
    author = {De Luca, V. and Franciolini, G. and Kehagias, A. and Peloso, M. and Riotto, A. and \"Unal, C.},
    title = "{The Ineludible non-Gaussianity of the Primordial Black Hole Abundance}",
    eprint = "1904.00970",
    archivePrefix = "arXiv",
    primaryClass = "astro-ph.CO",
    doi = "10.1088/1475-7516/2019/07/048",
    journal = "JCAP",
    volume = "07",
    pages = "048",
    year = "2019"
}

@article{Bardeen:1985tr,
    author = "Bardeen, James M. and Bond, J. R. and Kaiser, Nick and Szalay, A. S.",
    title = "{The Statistics of Peaks of Gaussian Random Fields}",
    reportNumber = "FERMILAB-PUB-85-148-A, NSF-ITP-85-93",
    doi = "10.1086/164143",
    journal = "Astrophys. J.",
    volume = "304",
    pages = "15--61",
    year = "1986"
}

@article{Iacovelli:2022mbg,
    author = "Iacovelli, Francesco and Mancarella, Michele and Foffa, Stefano and Maggiore, Michele",
    title = "{GWFAST: A Fisher Information Matrix Python Code for Third-generation Gravitational-wave Detectors}",
    eprint = "2207.06910",
    archivePrefix = "arXiv",
    primaryClass = "astro-ph.IM",
    doi = "10.3847/1538-4365/ac9129",
    journal = "Astrophys. J. Supp.",
    volume = "263",
    number = "1",
    pages = "2",
    year = "2022"
}
 
\end{document}